%% file: ex_article.tex
\documentclass[final,onefignum,onetabnum]{siamonline250106}

\input{ex_shared}

\input{macros}

\ifpdf
\hypersetup{
  pdftitle={A Multi-fidelity Estimator of the Expected Information Gain for Bayesian Optimal Experimental Design},
  pdfauthor={Thomas E. Coons, and Xun Huan}
}
\fi

\begin{document}

\maketitle

\begin{abstract}
Optimal experimental design (OED) is a framework that leverages a mathematical model of the experiment to identify optimal conditions for conducting the experiment. Under a Bayesian approach, the design objective function is typically chosen to be the expected information gain (EIG). However, EIG is intractable for nonlinear models and must be estimated numerically. Estimating the EIG generally entails some variant of Monte Carlo sampling, requiring repeated data model and likelihood evaluations---each involving solving the governing equations of the experimental physics---under different sample realizations. This computation becomes impractical for high-fidelity models. 
We introduce a novel multi-fidelity EIG (MF-EIG) estimator under the approximate control variate (ACV) framework.
This estimator is unbiased with respect to the high-fidelity mean, and minimizes variance under a given computational budget.
We achieve this by first reparameterizing the EIG so that its expectations are independent of the data models, a requirement for compatibility with ACV. We then provide specific examples under different data model forms, as well as practical enhancements of sample size optimization and sample reuse techniques.
We demonstrate the MF-EIG estimator in two numerical examples: a nonlinear benchmark and a turbulent flow problem involving the calibration of shear-stress transport turbulence closure model parameters within the Reynolds-averaged Navier--Stokes model. We validate the estimator's unbiasedness and observe one- to two-orders-of-magnitude variance reduction compared to existing single-fidelity EIG estimators. 
\end{abstract}



\section{Introduction}
\label{intro}

Experiments and the collection of empirical data are integral to the scientific process. Of particular interest in modern engineering applications is the use of experimental data to develop predictive models. However, conducting experiments and collecting data can be difficult, expensive, time-consuming, and even dangerous. Therefore, it is crucial to select the best experimental design based on an appropriate measure in order to maximize the experiment's value. 

Optimal experimental design (OED) (see, e.g.,~\cite{Huan2024} for a recent review) is a framework that leverages a mathematical model of the experiment to identify optimal design conditions.
Bayesian OED~\cite{chaloner_bayesian_1995,Ryan2016,Alexanderian2021,Rainforth2023,Strutz2024} further incorporates Bayesian concepts of prior and posterior distributions, enabling the use of information-theoretic criteria for guiding design decisions. 
The design objective function, often referred to as the expected utility, is typically chosen to be the expected information gain (EIG) in the model parameters, or equivalently, the mutual information between the parameters and observation data~\cite{lindley_measure_1956}.
While the EIG can be computed analytically for linear models with Gaussian conjugate priors, it is intractable for nonlinear models where the observation data depends nonlinearly on the model parameters, and therefore it must instead be estimated numerically. 

Approaches for estimating the EIG typically entail some variant of Monte Carlo (MC) sampling to ``discretize'' the expectations (integrals) involved, 
for example in the nested MC (NMC) estimator~\cite{ryan_estimating_2003}.
Consequently, EIG calculation requires repeated evaluations of the data model under different parameter realizations, either to generate samples of the observation data or to compute likelihood values. Each evaluation involves solving the governing equations of the underlying experimental physics---such as a system of differential equations---which is often computationally intensive. 
To address this challenge, various acceleration techniques have been proposed to accelerate EIG estimation. These include, among others, improving sampling efficiency through importance sampling~\cite{Beck2018,Feng2019,Englezou2022}, developing lower and upper bounds for the EIG~\cite{Barber2003,Poole2019,Foster2019,Kleinegesse2020}, simplifying calculations with Gaussian approximations~\cite{Long2013,Overstall2018}, and substituting the original model with surrogate or reduced-order models~\cite{huan_simulation-based_2013,Duong2023}.

The use of surrogate and reduced-order models in OED is especially appealing in physics-based modeling, where varying levels of simplifications can be introduced to achieve computational speedups. This approach naturally leads to a \emph{multi-fidelity ensemble} of models, with each model offering different trade-offs between accuracy and computational cost.
While simply replacing the original model with a low-fidelity model is straightforward, it introduces additional bias. This surrogate-induced bias, however,  can be avoided with carefully designed multi-fidelity methods~\cite{Eldred2017,peherstorfer_survey_2018, biehler_multifidelity_2019, geraci_multifidelity_2020}.
Among these methods, 
multi-fidelity MC (MFMC) \cite{peherstorfer_optimal_2016} stands out for its ability to improve MC efficiency by exploiting the correlations among available models, reducing MC estimator variance for a given computational budget.
The approximate control variate (ACV) framework \cite{gorodetsky_generalized_2020,bomarito_optimization_2022} further generalizes MFMC, offering even greater variance reduction by more effectively leveraging correlations within the multi-fidelity model ensemble.
A different MC-based formulation that also provides variance reduction and accommodates multi-fidelity model ensembles is the multilevel best linear unbiased estimator (MLBLUE) \cite{schaden_multilevel_2020}, whose connection to ACV is explored in the notion of grouped ACV~\cite{gorodetsky_grouped_2024}.

The application of ACV to OED problems remains relatively unexplored. The most closely related studies involve multilevel MC (MLMC)~\cite{giles_multilevel_2015}, which has also been shown to be a special case of the ACV family of estimators. In MLMC, EIG estimators are constructed where the level is controlled by the number of inner-loop samples in the NMC estimator~\cite{goda_multilevel_2020} or the discretization resolution of the governing differential equations~\cite{beck_multilevel_2020}. 
Importantly, MLMC 
relies on specific assumptions about the cost-accuracy relationships between models and theirconvergence towards a ``truth model'' in the limit.
In contrast, ACV offers greater flexibility, as it can accommodate any ensemble of models regardless of whether their cost-accuracy relationships are known. 
ACV thus can incorporate diverse and flexible mixes of models, including those derived from simplified physics, different numerical methods, or data-driven approximations. In comparison, the theory surrounding MLMC is limited to model ensembles that satisfy specific cost-accuracy and convergence assumptions, such as those based on coarsened discretizations or relaxed convergence criteria. 
Furthermore, ACV directly minimizes error towards the high-fidelity mean---with respect to this goal, MLMC is generally suboptimal compared to other ACV estimators.

In this paper, we introduce a novel multi-fidelity EIG (MF-EIG) estimator for OED problems under the ACV framework.
Notably, this estimator is unbiased with respect to the high-fidelity mean,
and minimizes variance under a given computational budget. 
Our approach can also be tailored to accommodate other expected utility choices beyond EIG; however, this work emphasizes the nuances and intricacies of the EIG formulation.
Our contributions are summarized as follows.
\begin{itemize}
\vspace{2mm}
\item We derive the MF-EIG estimator, which accommodates general multi-fidelity ensembles of models under the ACV framework. Implementation is facilitated using the Multi-Model Monte Carlo with Python (MXMCPy) toolbox~\cite{bomarito_multi_2020}, allowing a wide range of ACV allocation strategies that encompasses existing MFMC and MLMC methods as special cases.
Our code is available at: \url{https://github.com/tcoonsUM/mf-eig/}.
\vspace{2mm}
\item We present a reparameterization of the EIG that reformulates its expectations to be independent of the data models, a requirement for compatibility with the ACV framework. 
This reparameterization also provides insights into the inner mechanics of EIG estimation and the notion of data model fidelity. 
We provide specific examples of NMC-style utility models under additive independent
and scaled data noise forms, and propose
practical enhancements via inner-loop sample size optimization and sample reuse techniques.
\vspace{2mm}
\item We demonstrate the MF-EIG estimator in two numerical examples: a nonlinear benchmark and a turbulent flow problem involving the calibration of shear-stress transport turbulence closure model parameters within the Reynolds-averaged Navier--Stokes model. We validate the estimator's unbiasedness and 
variance reduction through these examples, and provide explanations to the behavior of results through physical principles and insights. 
\end{itemize}
\vspace{2mm}

The remainder of this paper is organized as follows. Section \ref{sec:prob-statement} introduces the OED problem formulation.
Section~\ref{sec:ACVs} reviews the ACV framework in a general context.  
Section \ref{sec:four} introduces the MF-EIG estimator, including the required reparameterization, examples for different data models, and properties and enhancement techniques for the estimator. 
Section \ref{sec:ex} evaluates the MF-EIG estimator through a nonlinear benchmark and a turbulent flow design problem, highlighting its performance in variance reduction.
Section \ref{sec:conclusions} concludes with key findings and directions for future work.

\section{Problem formulation}
\label{sec:prob-statement}

We first introduce the OED problem that we ultimately seek to solve. 

\subsection{OED problem statement}

Let $(\Omega, \mathcal{F}, \mathbb{P})$ be a probability space with sample space $\Omega$, $\sigma$-field $\mathcal{F}$, and probability measure $\mathbb{P}$ defined on $(\Omega, \mathcal{F})$. 
The unknown model parameters are treated as a random vector and denoted\footnote{We use upper case to denote a random variable or random vector, and lower case to denote its realization.} as $\Theta:\Omega \rightarrow \mathbb{R}^{n_{\theta}}$;
the design of the experiment (i.e., the experimental conditions that we can control) is denoted as $\design \in \Design \subseteq \mathbb{R}^{n_{\design}}$;
and the experiment's observation data is denoted as $Y \in \mathbb{R}^{n_{y}}$.
We further assume all distribution functions are absolutely continuous with respect to the Lebesgue measure such that probability density functions exist. 
Then, when an experiment is performed at $\design$ and yields an observation $Y = y$, the probability density of $\Theta$ is updated according to Bayes' rule:
\begin{align}
    p(\theta | y, \design) = \frac{p(y|\theta , \design) \, p(\theta)} {p(y|\design)},
\label{eqn:bayes}
\end{align}
where $p(\theta)$\footnote{Formally, each density function is identified through a subscript that indicates its corresponding random variable, e.g., $p_{\Theta}(\theta)$, $p_{Y|\theta,\design}(y|\theta,\design)$, $p_{Y|\design}(y|\design)$, and $p_{\Theta|y,\design}(\theta|y,\design)$. To alleviate the burden of notation, we omit the subscript when the same information is clear from the arguments of the density function. However, we will make the subscript explicit when it is helpful for clarification.} is the prior density that reflects the initial uncertainty on $\Theta$ (the prior's dependence on $\design$ is omitted since knowledge on $\Theta$ should not depend on the pending experimental design); $p(y|\theta, \design)$ is the likelihood; $p(y|\design)$ is the evidence or marginal likelihood; and $p(\theta | y, \design)$ is the posterior density that reflects the updated uncertainty on $\Theta$ after conditioning on the experimental outcome. 

Following a decision-theoretic formulation of the OED problem~\cite{lindley_measure_1956}, we establish an expected utility
\begin{align}
\label{eqn:EU}
    U(\design) := \EE_{Y,\Theta|\design} \left[u(\design,Y,\Theta)\right],
\end{align}
where the utility function $u(\design,y,\theta)$ reflects the value brought by an experiment and may depend on the design $\design$, the observation $y$, and the true parameter value $\theta$. Since $y$ is not known when designing the experiment and $\theta$ is uncertain, an expectation is taken jointly over $Y,\Theta |\design$ to account for their possible values. In particular, we select the posterior-to-prior log density ratio as the utility: 
\begin{align}
\label{eqn:u_logratio}
    u(\design,y,\theta) = \log\left[ \frac{p(\theta|y,\design)}{p(\theta)}\right] = \log\left[ \frac{p(y|\theta,\design)}{p(y|\design)}\right],
\end{align}
where the second equality results from a direct application of Bayes' rule in \eqref{eqn:bayes}, yielding the equivalent likelihood-to-evidence log density ratio. 
It has been shown~\cite{Bernardo1979} that in order for the utility function to be a smooth, proper, and local scoring rule~\cite{Gneiting2007}, the utility must follow such a logarithmic form.
Upon substituting \eqref{eqn:u_logratio} into \eqref{eqn:EU}, the expected utility becomes
\begin{align}
\label{eqn:EIG1}
       U(\design) =      
       \EE_{Y,\Theta|\design} \left[\log\left[ \frac{p(\Theta|Y,\design)}{p(\Theta)}\right]\right] = \int_{Y} \int_{\Theta} p(y,\theta|\design)\,\log\left[ \frac{p(\theta|y,\design)}{p(\theta)}\right] \, \text{d}\theta \, \text{d}y.
\end{align}
The resulting $U(\design)$ is the EIG in $\Theta$ at design $\design$, or equivalently, the mutual information between $\Theta$ and $Y$ at design $\design$.
This is a general formulation of the expected utility that does not require assumptions on linearity or Gaussianity.

Additional intuition on the expected utility in \eqref{eqn:EIG1} can be obtained by recognizing that the same $U(\design)$ can be achieved using an alternative utility based on the Kullback--Leibler (KL) divergence:
\begin{align}\label{eqn:kldiv}
u^{\text{div}}(\design,y,\theta) = D_{\text{KL}}\left(p_{\Theta|y,\design}\,||\,p_{\Theta}\right) 
= \int_{\widetilde{\Theta}} p(\tilde{\theta}|y,\design)\,\log\left[ \frac{p(\tilde{\theta}|y,\design)}{p(\tilde{\theta})}\right] \text{d}\tilde{\theta} =u^{\text{div}}(\design,y),
\end{align}
where $D_{\text{KL}}$ is the KL divergence operator, and we note that in the end $u^{\text{div}}(\design,y)$ happens to be independent of $\theta$ since the parameters are already integrated out by the KL divergence. One can verify the equivalence to \eqref{eqn:EIG1} as $\EE_{Y,\Theta|\design} \left[u^{\text{div}}(\design,y)\right]=\EE_{Y,\Theta|\design} \left[u(\design,y,\theta)\right]=U(\design)$. 
The interpretation of \eqref{eqn:kldiv} is that, a larger $u^{\text{div}}$ reflects a scenario with a more substantial posterior update relative to the prior distribution, and hence is more informative.

Finally, the OED problem entails finding an optimal design $\design^{\ast}$ that maximizes the expected utility:
\begin{align}\label{optimProb}
    \design^{\ast} \in \argmax_{\design \in \Design} U(\design) .
\end{align}

\subsection{Nested Monte Carlo estimator for the EIG}
\label{sec:NMC}

In general, the EIG expected utility in \eqref{eqn:EIG1} does not have an analytical form and must be estimated numerically. One popular approach is the NMC estimator~\cite{ryan_estimating_2003}, which we will adopt as a baseline for comparison:
\begin{align}
    U(\design) &=  
    \int_{Y} \int_{\Theta} p(y,\theta|\design) \, \log\left[ \frac{p(y|\theta,\design)}{p(y|\design)}\right]  \, \text{d}\theta \, \text{d}y\label{eqn:EIG2} \\
    &\approx \frac{1}{N_{\text{out}}} \sum^{N_{\text{out}}}_{i=1} \left\{ 
    \log\left[p(y^{(i)}|\theta^{(i)},\design)\right]-\log\left[p(y^{(i)}|\design)\right] 
    \right\} \label{eqn:outer}
    \\
    &\approx    \frac{1}{N_{\text{out}}} \sum^{N_{\text{out}}}_{i=1} \left\{ 
    \log\left[p(y^{(i)}|\theta^{(i)},\design)\right]-\log\left[ \frac{1}{N_{\text{in}}} \sum^{N_{\text{in}}}_{j=1} 
    p(y^{(i)}|\theta^{(i,j)},\design) \right] 
    \right\}
    := \hat{U}^{\text{NMC}}(\design), 
    \label{eqn:NMC}
\end{align}
where the initial equality results from transforming \eqref{eqn:EIG1} to its likelihood-to-evidence log density ratio version, and the approximation expressions follow from standard MC estimations; $\theta^{(i)}$ are independent and identically distributed (i.i.d.) samples from the prior $p(\theta)$, and $y^{(i)}$ are corresponding independent samples from the likelihood $p(y|\theta^{(i)},\design)$; and $N_{\text{out}}$ and $N_{\text{in}}$ are the number of MC samples in the outer and inner loops, respectively.
The complexity of the NMC estimator at a given design $\design$, in terms of likelihood operations, is $\mathcal{O}(N_{\text{out}}N_{\text{in}})$. The estimator is biased with respect to the exact EIG under finite $N_{\text{in}}$, but is asymptotically unbiased in the limit as $N_{\text{in}}\to \infty$.  
Detailed analysis of the NMC estimator can be found in~\cite{ryan_estimating_2003,Beck2018,rainforth2018nesting}. 
We note that alternative estimators---such as those referenced in Section~\ref{intro}---can also be employed; however, they may introduce additional bias, for example when relying on Gaussian approximations or surrogate models.

\section{Background on approximate control variates}
\label{sec:ACVs}

We now provide a general introduction to the ACV framework.
Consider a random vector $Z \in \mathbb{R}^{n_{z}}$ and a random variable $Q \in \mathbb{R}$ related by a mapping $Q=f(Z)$. We are interested in obtaining the expectation of $Q$, which can be approximated by a standard $N$-sample MC estimator:
\begin{align}\label{eqn:mc}
    \mathbb{E}\left[Q\right] = \mathbb{E}_{Z}\left[f(Z)\right] 
    \approx \hat{Q}(z) := \frac{1}{N} \sum^{N}_{j=1}f(z^{(j)}),
\end{align}
where $z^{(j)} \sim p(z)$ are i.i.d. samples drawn from the distribution of $Z$, and $z:=\left[ z^{(1)},\ldots,z^{(N)} \right]$ denotes the set of $N$ random samples. Oftentimes, we will emphasize the explicit dependence of the MC estimator on the realized sample values via the notation $\hat{Q}(z)$. 
The MC estimator $\hat{Q}$ is unbiased and has
variance $N^{-1}\Var[Q]$.

One technique to reduce the estimator variance
is through the linear control variate (CV) algorithm~\cite{lavenberg_statistical_1982}. 
We adopt the CV notation where a subscript 0 indicates the target (e.g., high-fidelity) quantity, i.e., $Q_0=Q$ and $f_0=f$, and we remain interested in obtaining the expectation of $Q_0=f_0(Z)$. 
Additionally, an ensemble of $M$ auxiliary random variables, $Q_{m}$ for $m=1,\ldots,M$,  are introduced resulting from corresponding (e.g., low-fidelity) models, $Q_m =f_{m}(Z)$. Furthermore, we assume their expected values $\mu_{m}:=\EE[Q_{m}]$ are known.
A CV estimator, $\hat{Q}^{\text{CV}}$, can then be formed by assembling MC estimates induced by the different models, $\hat{Q}_m(z) := \frac{1}{N} \sum^{N}_{j=1}f_m(z^{(j)})$,  
each evaluated at the \emph{same} set of $N$ random samples, $z$:
\begin{align}\label{eqn:OCV}
    \hat{Q}^{\text{CV}}(z;\alpha) := 
    \hat{Q}_{0}(z)+\sum^{M}_{m=1}\alpha_{m}\left( \hat{Q}_{m}(z)-\mu_{m} \right),
\end{align}
where $\alpha:=\left[\alpha_{1},...,\alpha_{M}\right]$ is the vector of scalar CV weights.
By taking expectation on both sides of \eqref{eqn:OCV}, one can verify that $\hat{Q}^{\text{CV}}$ is unbiased with respect to the high-fidelity mean for any $\alpha$, i.e., $\EE[\hat{Q}^{\text{CV}}]=\EE[Q_0]$. 
The optimal CV weight vector, $\alpha^{\text{CV}}$, that produces the minimum variance for $\hat{Q}^{\text{CV}}$ can be computed analytically if covariance information is available: let $C \in \mathbb{R}^{M\times M}$ denote the covariance matrix for the 
$Q_{m}$ variables, and $c \in \mathbb{R}^{M}$ the vector of covariances between each $Q_{m}$ and the target $Q_{0}$, one can show that $\alpha^{\text{CV}}=-{C}^{-1}{c}$. 

Directly applying the CV formula in \eqref{eqn:OCV} is generally not possible since $\mu_{m}$, the exact means for the auxiliary random variables, are typically not known. ACV methods~\cite{gorodetsky_generalized_2020, bomarito_optimization_2022} replace these exact means with estimated means. 
Let $\hat{\mu}_{m}$ denote the estimated mean for the $m$th model obtained via standard MC.
The ACV estimator, $\tilde{Q}$, then takes the form:
\begin{align}
        \tilde{Q}(z;\alpha,\mathcal{A}) :=& \,\,
        \hat{Q}_{0}(z_{0})+\sum^{M}_{m=1}\alpha_{m}\left( \hat{Q}_{m}(z^{\ast}_{m})-\hat{\mu}_{m}(z_{m}) \right) \label{eqn:ACV-Formula} \\
        =& \,\, \hat{Q}_{0}(z_{0})+\sum^{M}_{m=1}\alpha_{m}\left( \hat{Q}_{m}(z^{\ast}_{m})-\hat{Q}_{m}(z_{m}) \right),  \nonumber   
\end{align}
where now we have further allocated the input sample vector $z$ into subsets $z_{0}\neq \emptyset$, $z_{m}$, and  $z^{\ast}_{m}$ such that their union is $z$, and hence the MC estimators $\hat{Q}_m$ all have different sample sizes corresponding to their specific input arguments. There is no strict relationship required between these sample subsets, i.e., the intersections between subsets 
can be empty or non-empty. Information regarding the sample allocation, including which samples are shared among different subsets and the size of each subset, is encoded in $\mathcal{A}$.
This contrasts with CV, where an identical sample set is used across all $m$. 
Most tools also support the exclusion of specific models by setting their corresponding sample allocations to $\emptyset$, as is implemented in the `Automatic Model Selection' feature of MXMCPy \cite{bomarito_multi_2020}. This enables users to include all available models without a priori selection, with the only tradeoff being a slight increase in computation time when solving the optimal sample allocation.

ACV estimators can offer significant computational savings compared to standard MC, especially if the low-fidelity models are inexpensive to evaluate and well correlated to the high-fidelity model. 
Similar to CV, one can verify that $\tilde{Q}$ is unbiased, i.e., $\EE[\tilde{Q}]=\EE[Q_0]$, by taking the expectation on both sides of \eqref{eqn:ACV-Formula}. 
Since the ACV estimator is unbiased with respect to the high-fidelity mean, the mean square error (MSE) of the estimator is entirely the estimator variance. To easily express this variance, we first introduce a vectorized notation for the ACV estimator:
\begin{align}\label{eqn:vectorized-acv}
    \tilde{Q}(z;\alpha,\mathcal{A}) = \hat{Q}_{0} + \alpha^{\intercal}\Delta,
\end{align}
where $\Delta:=\left[ \Delta_{1}(z_{1}^{\ast},z_{1}), \ldots,  \Delta_{M}(z_{M}^{\ast},z_{M})\right]$ with $\Delta_{m}(z_{m}^{\ast},z_{m}):=\hat{Q}_{m}(z^{\ast}_{m})-\hat{Q}_{m}(z_{m})$, and the explicit dependence on the input samples $z_{m}$ and $z_{m}^{\ast}$ are omitted for simplicity. The optimal weight vector for a given ACV sample allocation $\mathcal{A}$ is then given by~\cite{bomarito_optimization_2022}:
\begin{align}\label{eqn:alpha-star-acv}
    \alpha^{\ast}(\mathcal{A})=-\Cov[\Delta, \Delta]^{-1}\Cov[\Delta, \hat{Q}_{0}],
\end{align}
where the covariance terms $\Cov[\Delta, \Delta]$ and $\Cov[\Delta, \hat{Q}_{0}]$ can be obtained analytically given $C$, the covariance matrix for the $Q_m$ variables. 
We note that $\alpha^{\ast}$ depends on $\mathcal{A}$ but we omit explicitly writing this dependence in the remainder of this paper. 
When the weights are set to these optimal values, the ACV estimator variance becomes~\cite{gorodetsky_generalized_2020}:
\begin{align}\label{eqn:acv-variance}
    \Var[\tilde{Q}(Z;\alpha^{\ast},\mathcal{A})] = \Var[\hat{Q}_{0}] - \Cov[\Delta, \hat{Q}_{0}]^{\intercal} \Cov[\Delta, \Delta]^{-1}\Cov[\Delta, \hat{Q}_{0}].
\end{align}

The sample allocation $\mathcal{A}$ determines how much computational workload is distributed to each $\hat{Q}_m$. In fact, both MFMC \cite{peherstorfer_optimal_2016} and MLMC \cite{giles_multilevel_2015} are special cases of ACV that differ in the family of possible $\mathcal{A}$ choices and how $\alpha$ is set. For example, MFMC restricts each $z_{m}^{\ast}=z_{m-1}$ and $z_{m} \subset z_{k}$ for $m<k$, and MLMC uses $\alpha_m=-1$ for all $m=1,\ldots,M$. We also note that while the MLMC estimator is in fact an ACV estimator by structure, the theory \cite{giles_multilevel_2015} regarding its error convergence and computational complexity bounds only apply under multilevel assumptions on the model ensemble used. In \cite{gorodetsky_generalized_2020}, the authors generalize these rules to enable additional families of ACV estimators, reaching 16 parametric families in \cite{bomarito_optimization_2022}. 
They further derive an expression for the variance of the ACV estimator, allowing one to compare and optimize across different allocations:
\begin{align}
    \min_{\mathcal{A}\in\mathbb{A}} &\quad \Var[\tilde{Q}(Z;\alpha^{\ast},\mathcal{A})] \label{eqn:mxmcpy}\\
    \text{subject to} &\quad \mathcal{W}(w,\mathcal{A})\leq w_{\text{budget}},\label{eqn:mxmcpy_constraint}
\end{align}
where $\mathbb{A}$ is the set of allowable sample allocations, $w_{\text{budget}}$ is the total computational budget, and $\mathcal{W}(w,\mathcal{A})$ computes the total cost of the estimator under the vector of model costs $w := [w_0, \ldots, w_M]$ and sample allocation $\mathcal{A}$.
In this work, we solve this optimization problem using the MXMCPy toolbox \cite{bomarito_multi_2020}, a Python implementation of the ACV estimator design approach described in \cite{bomarito_optimization_2022}. 

The optimal CV weight vector $\alpha^{\ast}$ in \eqref{eqn:alpha-star-acv} and the ACV estimator variance in \eqref{eqn:acv-variance} require the covariance information among the different models, which is usually unknown. A common approach, which we adopt in this work, is to estimate these covariance terms empirically from simple pre-allocated pilot samples shared across different models. 
Even under limited pilot sample sizes, the ACV method has been shown to achieve significant variance reduction~\cite{pham_ensemble_2022}. 
We use the pilot samples to also estimate the average computational cost for each model. 
We note that pilot sampling incurs an additional offline cost.
Advanced techniques using multi-arm bandit reinforcement learning have also been proposed~\cite{jakeman2022multi} that can more efficiently balance pilot sampling cost with estimator cost.

\section{A multi-fidelity estimator for the EIG}\label{sec:four}

We now apply the ACV framework introduced in Section~\ref{sec:ACVs} to the OED problem stated in Section~\ref{sec:prob-statement}.

\subsection{Reparameterization of the EIG}
\label{sec:alt-form}

The ACV framework assumes that the cost of generating input samples $z$ is negligible relative to the model evaluation costs, i.e., of $f_m(z)$.
However, this assumption does not hold when MC is directly applied to ``discretize'' the expectation of the expected utility in \eqref{eqn:EU}, such as for the NMC estimator in \eqref{eqn:NMC}. In these cases, sampling $Z=\{Y,\Theta\}$ involves drawing a sample of $Y$ from the likelihood, which requires 
evaluating the data model that simulates the experimental process. 
Mathematically, the data model can be written as $Y=h(\Noise; \Theta, \design)$, where $\Noise$ represents the random data noise that is assumed to be independent of $\Theta$, simple to sample, and whose probability density is easy to evaluate. 
A common example of $h$ is an additive independent noise model in the form $Y=g(\Theta,\design)+\Noise$, where $g$ is a deterministic forward model (e.g., governed by a system of differential equations) and $\Noise$ is a Gaussian random noise. 
However, we emphasize that the function $h$ in our formulation is general and not restricted to the additive noise case.
To satisfy the assumption for ACV, we perform a change of variables to the EIG such that the new variables of expectation can be sampled without requiring any data model.

We first reparameterize the likelihood. 
Consider cases where $h$ is injective and continuously invertible, and further defining $\Noise = h^{-1}(Y;\Theta,\design)$ as the inverse of $h$, then the likelihood can be rewritten as
\begin{align}\label{eqn:llh}
    p
    (y|\theta,\design) = p_{\Noise|\design}\left(h^{-1}( y; \theta,\design ) \big|\design \right) \, \big\lvert J^{-1}(y;\theta,\design) \big\rvert
    = p
    (\noise |\design) \, \big\lvert J^{-1}(h(\noise;\theta,\design);\theta,\design) \big\rvert,
\end{align}
where $\left\lvert J^{-1} \right\rvert=\left\lvert \frac{\partial h^{-1}}{\partial y} \right\rvert$ denotes the absolute value of the determinant of the Jacobian matrix of $h^{-1}$, $p_{\Noise|\design}$ may have a $\design$-dependence but has been earlier assumed to be independent of $\Theta$, and the last equality makes the substitution of $y=h(\noise;\theta,\design)$ and its inverse $\noise=h^{-1}(y;\theta,\design)$. 
Correspondingly, the evidence becomes
\begin{align}
    p(y|\design) = \int_{\widetilde{\Theta}}p(y|\tilde{\theta},\design)\, p(\tilde{\theta})\, \text{d}{\tilde{\theta}} =  \int_{\widetilde{\Theta}} p_{\Noise|\design}(\tilde{\noise} |\design) \,\big\lvert J^{-1}(y;\tilde{\theta},\design) \big\rvert \, p(\tilde{\theta}) \, \text{d}\tilde{\theta},\label{eqn:evi_eps}
\end{align}
where $\tilde{\noise} = h^{-1}(y;\tilde{\theta},\design)$. 
Since $y=h(\noise;\theta,\design)$, $\tilde{\noise}$ is also dependent on the ``true'' $\noise$ that, together with the ``true'' $\theta$, generated the $y$ at which $p(y|\design)$ is being evaluated.

Next, we apply \eqref{eqn:llh} and \eqref{eqn:evi_eps} to 
the EIG expression in \eqref{eqn:EIG2} to rewrite it as an integral over $\noise$ instead of $y$:
\begin{align}
    U(\design) 
    &=
    \int_{Y} \int_{\Theta} p(\theta) \, p(y|\theta,\design) \log\left[ \frac{p(y|\theta,\design)}{p(y|\design)}\right] \, \text{d}\theta \, \text{d}y \nonumber\\ 
    &= 
    \int_{\Noise} \int_{\Theta} p(\theta)\, p(\noise |\design) \,  \big\lvert J^{-1}(h(\noise;\theta,\design);\theta,\design) \big\rvert \, \nonumber  \\ 
    & \hspace{6.5em} \times
    \log\left[ \frac{p(\noise |\design) \, \big\lvert J^{-1}(h(\noise;\theta,\design);\theta,\design) \big\rvert}{
    \int_{\widetilde{\Theta}} p_{\Noise|\design}(\tilde{\noise} |\design) \,\big\lvert J^{-1}(h({\noise};{\theta},\design);\tilde{\theta},\design) \big\rvert \, p(\tilde{\theta}) \, \text{d}\tilde{\theta}}\right]  \, \text{d}\theta \, \big\lvert J(\noise;\theta,\design) \big\rvert \, \text{d}\noise \nonumber\\ 
    &=
    \int_{\Noise} \int_{\Theta} p(\noise,\theta |\design) \,   \log\left[ \frac{p(\noise |\design) \, \big\lvert J^{-1}(h(\noise;\theta,\design);\theta,\design) \big\rvert}{
    \int_{\widetilde{\Theta}} p_{\Noise|\design}(\tilde{\noise} |\design) \,\big\lvert J^{-1}(h(\noise;\theta,\design);\tilde{\theta},\design) \big\rvert \, p(\tilde{\theta}) \, \text{d}\tilde{\theta}}\right] \, \text{d}\theta \, \text{d}\noise \label{eqn:newEIG_int}\\
    &=\EE_{\Noise,\Theta|\,\design} \left[\log\left[ \frac{p(\Noise |\design) \, \big\lvert J^{-1}(h(\Noise;\Theta,\design);\Theta,\design) \big\rvert}{\int_{\widetilde{\Theta}} p_{\Noise|\design}(\tilde{\Noise} |\design) \,\big\lvert J^{-1}(h({\Noise};{\Theta},\design);\tilde{\theta},\design) \big\rvert \, p(\tilde{\theta}) \, \text{d}\tilde{\theta}}\right]\right] 
    = \EE_{\Noise,\Theta|\,\design} \left[u_{\Noise}(\xi,\Noise,\Theta)\right],
    \label{eqn:newEIG_ex}
\end{align}
where 
the second equality uses \eqref{eqn:llh} and \eqref{eqn:evi_eps} and substitutes $\text{d}y = \big\lvert J(\noise;\theta,\design) \big\rvert \, \text{d}\noise$, the third equality makes simplifications $\big\lvert J^{-1}(h(\noise;\theta,\design);\theta,\design) \big\rvert \, \big\lvert J(\noise;\theta,\design) \big\rvert=1$ and $p(\theta) p(\noise |\design)=p(\noise,\theta |\design)$ due to the assumed independence between $\Theta$ and $\Noise$, and the last equality introduces \begin{align}
    u_{\Noise}(\xi,\noise,\theta) :=
    \log\left[ \frac{p(\noise |\design) \, \big\lvert J^{-1}(h(\noise;\theta,\design);\theta,\design) \big\rvert}{\int_{\widetilde{\Theta}} p_{\Noise|\design}(\tilde{\noise} |\design) \,\big\lvert J^{-1}(h({\noise};{\theta},\design);\tilde{\theta},\design) \big\rvert \, p(\tilde{\theta}) \, \text{d}\tilde{\theta}}\right]
    \label{e:u_Noise}
\end{align}
to represent the now transformed utility inside the expectation.
We remind that $\tilde{\noise}$ is a function of $\noise$ since $\tilde{\noise} = h^{-1}(y;\tilde{\theta},\design)=h^{-1}(h(\noise;\theta,\design);\tilde{\theta},\design)$, and similarly $\tilde{\Noise}$ is a function of $\Noise$.
In the form of \eqref{eqn:newEIG_int} and \eqref{eqn:newEIG_ex}, the expectation (integration) in the EIG expression is now over $\Noise$ and $\Theta$ instead of $Y$ and $\Theta$, and consequently drawing samples of $Z=\{\Noise,\Theta\}$ no longer requires any data model.

\subsection{The MF-EIG estimator}
\label{sec:mf-eig}

With the EIG expressed as an expectation over $Z = \{\Noise,\Theta\}$ in \eqref{eqn:newEIG_ex}, 
we are now in position to form an ACV estimator for the EIG.
First, we introduce utility models, $u_m$ for $m=0,\ldots,M$, that given an input sample $z^{(i)}$, approximate the exact utility function:
\begin{align}\label{eqn:utility-models}
    u_{m}(\design,z^{(i)})\approx 
    u_{\Noise}(\xi,z^{(i)}),
    \quad m =0,\ldots,M.
\end{align}
Following the ACV convention from Section~\ref{sec:ACVs}, 
$u_{0}$ denotes the high-fidelity target utility model. However, $u_0$ will also be an approximation to $u_{\Noise}$ due to the intractable evidence term in the denominator of $u_{\Noise}$ that needs to be numerically estimated; we will discuss its implications in Section~\ref{sec:props}.
%
Then, applying the ACV formula in \eqref{eqn:ACV-Formula} to \eqref{eqn:newEIG_ex}, we obtain a 
MF-EIG estimator:
\begin{align}\label{eqn:MF-EIG}
    \tilde{U}(\design, z; \alpha, \mathcal{A}) := \hat{U}_{0}\left(\design,z_{0}\right)+\sum^{M}_{m=1}\alpha_{m}\left( \hat{U}_{m}\left(\design,z_{m}^{\ast}\right)-\hat{U}_{m}\left(\design,z_{m}\right) \right),
\end{align}
where each $\hat{U}_{m}$ represents the standard MC estimator of the EIG using $u_m$:
\begin{align}
    \hat{U}_{m}(\design,z_{m}) := \frac{1}{N_m} \sum_{i=1}^{N_m}{u}_{m}(\design,z_{m}^{(i)}),
\end{align} 
with $z_{m} = \left[ z^{(1)},\ldots,z^{(N_m)} \right] = \left[ \{\noise^{(1)},\theta^{(1)}\},\dots,\{\noise^{(N_m)},\theta^{(N_m)}\} \right]$ being the set of $N_m$ samples drawn from 
$p(\noise,\theta |\design)$
and allocated to $\hat{U}_{m}$ according to $\mathcal{A}$.

The utility models $u_m$ in \eqref{eqn:utility-models} are quite general, allowing flexibility in their definition to incorporate the numerical techniques chosen to estimate the exact utility and the quality of those estimates. For instance, all $u_m$ models might retain the same form as $u_{\Noise}$ and employ an inner-loop MC to estimate the evidence term, while utilizing  different 
data models
of varying fidelity. Alternatively, $u_m$ could all use the same 
data model
but vary in MC sample size, or apply a mix of techniques, such as some using MC, others relying on quadrature, and others leveraging density ratio estimation. 
Moreover, $u_m$ could even take a completely different form that does not converge to $u_{\Noise}$ while still considered as a low-fidelity estimate. 
Later in the paper, we will illustrate the MF-EIG framework by focusing on the first example.

\subsection{MF-EIG hyperparameters}

As outlined in Section \ref{sec:ACVs}, the variance-optimal values of ACV hyperparameters $\alpha$ and $\mathcal{A}$ depend on the covariance 
matrix $C$ for the vector of utility model outputs, denoted as $\mathrm{u}(\design,Z):=[u_0(\design,Z),\ldots,u_M(\design,Z)]^{\intercal}$,
and the computational costs of utility models. 
$C$ can be estimated by evaluating the models on an independent set of pilot samples of $Z$ and applying the unbiased sample covariance formula:
\begin{align}\label{eqn:cov}
    C&=\Cov\left[ 
    \mathrm{u}(\design,Z) 
    \right] \\
    &\approx \hat{\Sigma}(\design, z_{\text{pilot}}) := \frac{1}{N_{\text{pilot}}-1}\sum_{j=1}^{N_{\text{pilot}}}\left( \mathrm{u}(\design,z^{(j)}) - \overline{\mathrm{u}}(\design,z_{\text{pilot}}) \right) \left( \mathrm{u}(\design,z^{(j)}) - \overline{\mathrm{u}}(\design,z_{\text{pilot}}) \right)^{\intercal}, \nonumber
\end{align}
\sloppy where $N_{\text{pilot}}$ denotes the number of pilot samples in $z_{\text{pilot}}$, $\mathrm{u}(\design,z^{(j)})=[ u_{0}(\design,z^{(j)}),\ldots,u_{M}(\design,z^{(j)}) ]^{\intercal}$, and $\overline{\mathrm{u}}(\design,z_{\text{pilot}}):=[ \overline{u}_{0}(\design,z_{\text{pilot}}),\ldots,\overline{u}_{M}(\design,z_{\text{pilot}}) ]^{\intercal}$ with $\overline{u}_{m}(\design,z_{\text{pilot}}):=\frac{1}{N_{\text{pilot}}}\sum_{j=1}^{N_{\text{pilot}}}u_{m}(\design,z^{(j)})$. Using too few pilot samples can lead to an inaccurate or singular sample covariance matrix. To address this, we select a sufficiently large number of pilot samples in this work, ensuring $N_{\text{pilot}} \gg M$. To avoid repeated
pilot evaluations
at every $\design$ encountered, we adopt a single set of MF-EIG hyperparameters across all designs based on a design-averaged covariance matrix: 
\begin{align}\label{eqn:avg-cov}
    \overline{\Sigma}:=
    \frac{1}{N_{\design}}\sum_{k=1}^{N_{\design}}\hat{\Sigma}(\design_{k}),
\end{align}
where 
$\design_k$ are 
$N_{\design}$ designs selected for averaging (e.g., sampled uniformly in $\Design$). Additionally, the pilot samples can be used to estimate the average computational costs (e.g., CPU-hours of run time) of each utility model, if not already known. Once the covariance matrix and model costs are determined, the optimization problem in \eqref{eqn:mxmcpy} and \eqref{eqn:mxmcpy_constraint} is solved. The resulting hyperparameter values are then applied to all MF-EIG estimators across the design domain.
Adopting a single MF-EIG estimator structure and using uniform pilot sample sizes across the entire design domain is practical but may be suboptimal when the covariance matrix varies significantly throughout the domain. A more tailored approach could involved redesigning the MF-EIG estimator and allocating more or fewer pilot samples at different design locations, potentially increasing the total number of pilot samples needed to accurately estimate local covariance matrix. Alternatively, applying smoothing or surrogate modeling techniques to the covariance matrix across the design space may reduce the need for additional pilot samples. We leave the exploration of these strategies to future work.

\subsection{NMC-style utility models}
\label{sec:lf-mods}

We now provide a detailed presentation of MF-EIG estimators based on NMC-style utility models.
Similar to before, subscript $0$ denotes the high-fidelity target quantities introduced in Sections~\ref{sec:prob-statement} and \ref{sec:alt-form}; that is, $Y_0=Y$, $h_0=h$, $J_0=J$, 
$p_{Y_0|\theta,\design}(y_0|\theta,\design)=p_{Y|\theta,\design}(y|\theta,\design)$, $p_{Y_0|\design}(y_0|\design)=p_{Y|\design}(y|\design)$, and $u_{\Noise,0}=u_{\Noise}$.
Next, for $m=1,\ldots,M$, we introduce low-fidelity injective and continuously invertible data models $h_m$ relating the common random data noise $\Noise$ to low-fidelity observation variables $Y_m$, where $Y_m = h_m(\Noise;\Theta,\design)$ with inverse $\Noise = h_m^{-1}(Y_m;\Theta,\design)$ and Jacobian $J_m=\frac{\partial h_m}{\partial \noise}$ with inverse $J_m^{-1}=\frac{\partial h_m^{-1}}{\partial y_m}$. 
Each $Y_m$ is also associated with its own low-fidelity likelihood and evidence, $p(y_m|\theta,\design)$ and $p(y_m|\design)$.
Similar to \eqref{eqn:llh} and \eqref{eqn:evi_eps}, they can be transformed into
\begin{align}\label{eqn:llh2}
    p(y_m|\theta,\design) &= p_{\Noise|\design}\left(h_m^{-1}( y_m; \theta,\design ) \big|\design \right) \, \big\lvert J_m^{-1}(y_m;\theta,\design) \big\rvert
    = p
    (\noise |\design) \, \big\lvert J_m^{-1}(h_m(\noise;\theta,\design);\theta,\design) \big\rvert, \\[6pt] 
    p(y_m|\design) &= \int_{\widetilde{\Theta}}p(y_m|\tilde{\theta},\design)\, p(\tilde{\theta})\, \text{d}{\tilde{\theta}} =  \int_{\widetilde{\Theta}} p_{\Noise|\design}(\tilde{\noise}_m |\design) \,\big\lvert J_m^{-1}(y_m;\tilde{\theta},\design) \big\rvert \, p(\tilde{\theta}) \, \text{d}\tilde{\theta},\label{eqn:evi_eps2}
\end{align}
where $\tilde{\noise}_m = h_m^{-1}(y_m;\tilde{\theta},\design)$. 
Using $p(y_m|\theta,\design)$ and $p(y_m|\design)$, we can then form corresponding low-fidelity EIGs, and follow analogous derivations of \eqref{eqn:newEIG_int} and \eqref{eqn:newEIG_ex} to arrive at their transformed utility functions:
\begin{align}
    u_{\Noise,m}(\xi,\noise^{(i)},\theta^{(i)}) := \log\left[ \frac{p(\noise^{(i)} |\design) \, \big\lvert J_m^{-1}(h_m(\noise^{(i)};\theta^{(i)},\design);\theta^{(i)},\design) \big\rvert}{
    \int_{\widetilde{\Theta}} p_{\Noise|\design}(\tilde{\noise}_m^{(i)} |\design) \,\big\lvert J_m^{-1}(h_m({\noise}^{(i)};{\theta}^{(i)},\design);\tilde{\theta},\design) \big\rvert \, p(\tilde{\theta}) \, \text{d}\tilde{\theta}}\right].
\end{align}
Finally, upon estimating the denominators with inner-loop MC for all $u_{\Noise,m}$ terms for $m=0,\ldots,M$, we obtain the final set of NMC-style utility models:
\begin{align}\label{eqn:NMC-style}
    u_{m}(\design,z^{(i)}) := 
    \log{\left[ \frac{p(\noise^{(i)} |\design) \, \big\lvert J_{m}^{-1}(h_{m}(\noise^{(i)};\theta^{(i)},\design);\theta^{(i)},\design) \big\rvert} 
    { \frac{1}{N_{\text{in},m}} \sum_{j=1}^{N_{\text{in},m}} 
    p_{\Noise|\design}( \tilde{\noise}_m^{(i)} | \design ) \, \big\lvert J_{m}^{-1}(h_{m}({\noise}^{(i)};\theta^{(i)},\design);\tilde{\theta}^{(i,j)},\design) \big\rvert } \right]},
\end{align} 
for $m=0,\ldots,M$, and where $\tilde{\theta}^{(i,j)}$ are $N_{\text{in},m}$ i.i.d. samples from the prior $p(\theta)$.
In summary, $u_m$ for $m=0,\ldots,M$ are respective approximations to $u_{\Noise,m}$ resulting from the inner-loop MC, and  $u_{\Noise,m}$ for $m=1,\ldots,M$ are all approximations to the exact utility $u_{\Noise,0}=u_{\Noise}$ due to their low-fidelity data models $h_m$.

In the following, as examples, we present utility models for data models with additive independent and scaled data noise.

\subsubsection{Additive independent noise}
\label{sss:additive}

Data models with additive independent noise have the form
$Y_m=h_{m}(\Noise;\Theta,\design)=g_{m}(\Theta,\design)+\Noise$ with inverse $\Noise=h_{m}^{-1}(Y_m;\Theta,\design)=Y_m-g_{m}(\Theta,\design)$, where $g_m$ is the deterministic forward model, and $\Noise$ may depend on $\design$ but is assumed to be independent of $\Theta$ and the output of $g_{m}$. It is easy to see that $J_m$ and $J_{m}^{-1}$ are all identity matrices, with determinants equal to 1. 
The utility models in \eqref{eqn:NMC-style} subsequently become:
\begin{align}\label{eqn:additivelog}
    u_{m}(\design,z^{(i)}) &= 
    \log{\left[ \frac{p(\noise^{(i)} |\design)} 
    { \frac{1}{N_{\text{in},m}} \sum_{j=1}^{N_{\text{in},m}} 
    p_{\Noise|\design}( \tilde{\noise}_m^{(i)} | \design )} \right]} \\
    &= \log{\left[ \frac{p(\noise^{(i)} |\design)} 
    { \frac{1}{N_{\text{in},m}} \sum_{j=1}^{N_{\text{in},m}} 
    p_{\Noise|\design}\left( g_{m}(\theta^{(i)},\design) +\noise^{(i)} -  g_{m}(\tilde{\theta}^{(i,j)},\design) \big| \design \right)} \right]}  \nonumber
\end{align}
for $m=0,\ldots,M$.
The computational costs of evaluating the utility models are $w_{m}=(N_{\text{in},m}+1) \, w_{m}^{\,g}$, where $w_{m}^{\,g}$ is the cost of a single evaluation of the forward model $g_{m}$.

\subsubsection{Scaled noise}
\label{sss:scaled}

Data models with scaled noise have the form $Y_m=h_m(\Noise;\Theta,\design)=g_m(\Theta,\design)+g_m(\Theta,\design)\circ\Noise$
with inverse $\Noise=h_m^{-1}(Y_m;\Theta,\design)=(Y_m-g_m(\Theta,\design))\oslash g_m(\Theta,\design)$, where $\circ$ and $\oslash$ respectively denote element-wise (Hadamard) multiplication and division, and $\Noise$ again may depend on $\design$ but is assumed to be independent of $\Theta$ and the output of $g_{m}$. Such structure arises when the overall noise has a relative level compared to the signal $g_m$. For example, a ``10\% data noise'' can be modeled through $\mathcal{E}\sim \mathcal{N}(0,\text{diag}(0.1^2))$ under this data model form.
The Jacobian of $h_m^{-1}$ now depends on the model output, with
$J^{-1}_{m}(h_m(\noise;\theta,\design);\theta,\design) = \text{diag} (1 \oslash g_{m}(\theta,\design))$ and determinant
$\big\lvert J^{-1}_{m}(h_m(\noise;\theta,\design);\theta,\design) \big\rvert = 1 / \lvert \text{diag} ( g_{m}(\theta,\design))\rvert$,
where $\lvert  \text{diag} ( g_{m}(\theta,\design))\rvert$ is simply the product of the elements of the vector $g_{m}(\theta,\design)$. 
The utility models in \eqref{eqn:NMC-style} subsequently become:
\begin{align}
    u_{m}(\design,z^{(i)}) &= 
    \log \left[ \frac{ {p(\noise^{(i)} |\design)} \, \Big/ \,\left\lvert \text{diag} ( g_{m}(\theta^{(i)},\design))\right\rvert } 
    { \frac{1}{N_{\text{in},m}} \sum_{j=1}^{N_{\text{in},m}} 
     p_{\Noise|\design}( \tilde{\noise}^{(i)}_m | \design) \, \Big/ \, \left\lvert \text{diag} ( g_{m}(\tilde{\theta}^{(i,j)},\design))\right\rvert }  \right] \nonumber\\ 
    &= \log \left[ \frac{ \frac{
    \displaystyle
    {p(\noise^{(i)}|\design)}} {
    \displaystyle
    \left\lvert \text{diag} ( g_{m}(\theta^{(i)},\design))\right\rvert}} 
    {  
    \displaystyle
    \frac{1}{N_{\text{in},m}} \sum_{j=1}^{N_{\text{in},m}}  
     \frac{p_{\Noise|\design}\left( (g_{m}(\theta^{(i)},\design) +  g_{m}(\theta^{(i)},\design) \circ \noise^{(i)} - g(\tilde{\theta}^{(i,j)}, \design) ) \oslash g(\tilde{\theta}^{(i,j)}, \design) 
     \Big
     |\design \right)} { 
     \displaystyle 
     \left\lvert\text{diag} ( g_{m}(\tilde{\theta}^{(i,j)},\design))\right\rvert}} 
     \right]
\end{align}
for $m=0,\ldots,M$.
The computational costs of evaluating the utility models remain $w_{m}=(N_{\text{in},m}+1) \, w_{m}^{\,g}$.

\subsection{Properties of the MF-EIG estimator}\label{sec:props}

We now discuss properties of the MF-EIG estimator and propose practical techniques to enhance its performance. 

\subsubsection{Estimator bias} 
\label{ss:bias}

Inheriting the ACV properties in Section~\ref{sec:ACVs}, the MF-EIG estimator in \eqref{eqn:MF-EIG} is unbiased with respect to the high-fidelity mean, i.e., $\mathbb{E}_Z[\tilde{U}(\design, Z; \alpha, \mathcal{A})] = \mathbb{E}_Z[u_{0}(\design,Z)]$ for all $\design$, $\alpha$, and $\mathcal{A}$. 
However, $\mathbb{E}_Z[\tilde{U}(\design, Z; \alpha, \mathcal{A})] =\mathbb{E}_Z[u_{0}(\design,Z)]\neq \mathbb{E}_{\Noise,\Theta|\design}[u_{\Noise}(\design,\Noise,\Theta)]=U(\design)$ since $u_0$ generally is a numerical approximation to the exact utility $u_{\Noise}$. This $U$-bias, however, diminishes if $u_0 \to u_{\Noise}$ (e.g., as $N_{\text{in},0} \to \infty$ for the NMC-style $u_0$ in \eqref{eqn:NMC-style}); hence, MF-EIG is asymptotically unbiased for convergent $u_0$. 
Furthermore, the MSE of the MF-EIG estimator relative to $U$ can be decomposed as the sum of $U$-bias squared plus the estimator variance. The former is dictated by the available high-fidelity data model, the latter is the subject being minimized by the ACV framework via 
\eqref{eqn:mxmcpy}--\eqref{eqn:mxmcpy_constraint}.

The expectation of the MF-EIG estimator based on the NMC-style $u_{0}$ is the same as the expectation of the NMC estimator when they have the same (finite) number of inner-loop samples: $\mathbb{E}_Z[\tilde{U}(\design, Z; \alpha, \mathcal{A})]=\mathbb{E}_Z[u_{0}(\design,Z)]=\mathbb{E}[\hat{U}^{\text{NMC}}(\design)]$; hence, $\tilde{U}$ and $\hat{U}^{\text{NMC}}$ have the same $U$-bias. 
Properties of the NMC estimator have been extensively studied~\cite{ryan_estimating_2003,Beck2018,rainforth2018nesting}. 
Under finite $N_{\text{in}}$, the bias of $\hat{U}^{\text{NMC}}$ with respect to $U$ has leading term $A(\design) / N_{\text{in}}$, where $A(\design)$ depends on the design and distributions involved.
Furthermore, $A(\design)$ is always positive~\cite[Proposition 1]{Beck2018}, making the NMC estimator positively biased to leading order.
The bias goes to zero as $N_{\text{in}}\to \infty$, and thus $\hat{U}^{\text{NMC}}$ is an asymptotically unbiased and consistent (i.e., convergent in probability) estimator.
The MF-EIG estimator, when adopting a NMC-style $u_0$, then also has a $U$-bias with leading term $A(\design) / N_{\text{in},0} > 0$, and is asymptotically unbiased and consistent.
Ryan \cite{ryan_estimating_2003} further suggested that the NMC estimator bias may be approximately constant across the design domain $\Design$, which may be helpful for comparing the EIG across different $\design$.
We recommend keeping $N_{\text{in},0}$ reasonably large (within the budget constraints of the given application) so as to keep this bias in check. 

\subsubsection{Inner-loop sample sizes}

Hyperparameters of low-fidelity utility models
can be tuned by minimizing the overall ACV estimator variance~\cite{thompson_strategies_2023, bomarito_improving_2022}.
We take such an approach to optimize the inner-loop sample sizes, $N_{\text{in},m}$, for low-fidelity NMC-style utility models by building upon \eqref{eqn:mxmcpy}--\eqref{eqn:mxmcpy_constraint}:
\begin{align}
    \min_{\mathrm{N}_{\text{in}} \in \mathbb{N}^{M}} \quad \left\{ \quad 
    \begin{aligned}
    \min_{\mathcal{A}\in\mathbb{A}} &\quad \Var[\tilde{U}(\design,Z;\alpha^{\ast},\mathcal{A}, \mathrm{N}_{\text{in}})] \\
    \text{subject to} &\quad \mathcal{W}(w{(\mathrm{N}_{\text{in}}}),\mathcal{A})\leq w_{\text{budget}}
    \end{aligned}
    \quad \right\},
    \label{eqn:opt-nin}
\end{align}
where $\mathrm{N}_{\text{in}} = \left[ N_{\text{in},1}, \ldots,  N_{\text{in},M}\right]$ is the vector of low-fidelity inner-loop sample sizes, 
and the dependence of $\tilde{U}$ and $w$ on $\mathrm{N}_{\text{in}}$ are made explicit. 
We note that the high-fidelity inner-loop sample size, $N_{\text{in},0}$, is not included for optimization. This is because $N_{\text{in},0}$ primarily controls the bias of the MF-EIG estimator (see Section~\ref{ss:bias}), whose effect is not captured in the variance minimization goal of the ACV framework as seen in~\eqref{eqn:opt-nin}, and is therefore handled separately. 
The nested optimization problem in~\eqref{eqn:opt-nin} can be difficult to solve since the values of $\mathrm{N}_{\text{in}}$ are constrained to positive integers and the search domain of $\mathrm{N}_{\text{in}}$ grows as the number of low-fidelity models increases. Relaxing the integer problem and rounding its solution after completion is one possible way of alleviating these issues.

It is also important to note that the inner loop can suffer from arithmetic underflow when computed from small sample sizes, which further necessitates reasonably large values for $N_{\text{in},m}$ across utility models. Situations where $N_{\text{in},m}$ must remain small may benefit from enhancements such as importance sampling~\cite{Beck2018,Feng2019,Englezou2022}, and we note once again that the MF-EIG estimator can accommodate any such utility models. 

\subsubsection{Sample reuse in the inner loop}\label{secn:reuse}

Sampling the data model and evaluating the likelihood are typically the most expensive computational steps. For data models that can be expressed in terms of a deterministic forward model, $g$, such as the cases in Sections~\ref{sss:additive} and~\ref{sss:scaled}, the most costly operation is the evaluation of $g$ (e.g., solving a system of governing differential equations). 
One approach to reduce these costs in the NMC estimator is to reuse samples of ${\Theta}$ from the outer loop for the inner loop, so that the same  $g(\theta,\design)$ values can be reused without needing to recompute it on new $\theta$ samples. For example, Huan and Marzouk \cite{huan_simulation-based_2013} proposed setting the inner-loop and outer-loop samples to be identical sets, which requires $N_{\text{in}}=N_{\text{out}}$ and reduces the estimator complexity from $\mathcal{O}(N_{\text{out}}N_{\text{in}})$ to $\mathcal{O}(N_{\text{in}})$. 
While the sample-reuse strategy relieves computational burden, however, it introduces some additional bias to the NMC estimator. 

Directly reusing inner- and outer-loop samples for the MF-EIG estimator is not straightforward since the number of outer-loop samples for each utility model is not known until after solving the optimization problem in \eqref{eqn:mxmcpy}--\eqref{eqn:mxmcpy_constraint}, which requires the inner-loop sample sizes to be already decided.
However, some aspects of the sample-reuse strategy can still be injected by using the same inner-loop  samples of ${\Theta}$ \emph{across utility models}. 
Although this approach may not reduce computational costs---since evaluations of the forward model cannot be reused across utility models if they use different forward models---it can enhance the correlation among utility models, which is a central theme to the ACV framework. 

\section{Numerical examples}
\label{sec:ex}

We demonstrate the MF-EIG estimator in two numerical examples: Case 1---a one-dimensional nonlinear benchmark where we make extensive comparisons under different settings; and Case 2---a multi-dimensional turbulent flow problem where we also provide explanations to the behavior of results through physical principles and insights.

\subsection{Case 1: nonlinear benchmark}
\label{sec:toy}

We adopt the one-dimensional nonlinear data model introduced in~\cite{huan_simulation-based_2013}:
\begin{align}\label{eqn:toy}
    Y_0 = g_{0}(\Theta,\design) + \Noise
    = \Theta^{3}\design^{2} + \Theta \exp{( -|0.2-\design| )} +\Noise,
\end{align}
where $Y_0$, $\Theta$, $\design$, and $\Noise$ are scalars, parameter prior is $\Theta \sim \mathcal{U}(0,1)$, design is constrained to $\design\in\Design = [0,1]$, and data noise is Gaussian with $\Noise \sim \mathcal{N}(0,(10^{-2})^2)$. 
We further introduce low-fidelity data models based on low-fidelity forward models that approximate $g_{0}$ with degraded accuracy in the first term:
\begin{align}
   Y_1 &= g_{1}(\Theta,\design) + \Noise = k_{1}\,\Theta^{2.5}\design^{1.75} + \Theta \exp{( -|0.2-\design| )} + \Noise, \label{eqn:toy_y1}\\
   Y_2 &= g_{2}(\Theta,\design) + \Noise = k_{2}\,\Theta^{2}\design^{1.5} + \Theta \exp{( -|0.2-\design| )} + \Noise, \label{eqn:toy_y2}
\end{align}
where the constants $k_{1}= 0.5^{0.5}$ and $k_{2}= 0.5$ scale the first term of each model. 
To estimate the design-averaged utility model covariance matrix in \eqref{eqn:avg-cov}, we select $N_{\design}=41$ evenly-spaced design locations across $\Design$, each employing $N_{\text{pilot}}=500$ pilot samples to estimate the sample covariance following \eqref{eqn:cov}. 
For illustrative purposes, the high-fidelity forward model $g_{0}$ is assigned unit cost, $w^g_0=1$, and  the cost of evaluating $g_{1}$ and $g_{2}$ are set to $w^g_1=0.1$ and $w^g_2=0.01$, respectively\footnote{Despite artificially enforcing the model costs, 
the cost-correlation relationships of these toy models are very realistic if not slightly pessimistic in the sense that many well-designed low-fidelity models from coarsened grids, simplified physics, or data-driven techniques are better correlated for these costs, or cheaper for this level of correlation.}. 
All MF-EIG estimators are then formed from a total computational budget of $w_{\text{budget}} = 2.5 \times 10^{6}$.
Furthermore, we fix the inner-loop sample size of the high-fidelity utility model, $u_0$, to $N_{\text{in},0}=2500$. 
When applying the same total budget and inner-loop sample size to the single-fidelity NMC, it allows an outer-loop sample size of $N_{\text{out}} = 1000$ (after rounding up). We use the single-fidelity NMC estimator as a baseline comparison.  

Through the results, we highlight
and provide guidance on two aspects of the MF-EIG estimator: selection of the low-fidelity inner-loop sample sizes in Section~\ref{sec:innerdem}, and sample reuse in the inner loop in Section~\ref{sec:reuse_toy}. 
Lastly, we present a scaled noise variant of this problem in Section~\ref{sss:nonlinear_scaled}.

\subsubsection{Inner-loop sample sizes}\label{sec:innerdem}

We consider two approaches for determining the inner-loop sample sizes (no sample reuse). In the first approach, the inner-loop sample sizes are na\"{i}vely chosen to match that of the high-fidelity model, i.e., $N_{\text{in},0}=N_{\text{in},1}=N_{\text{in},2}=2500$. 
Any improvement in the estimator's performance thus arises solely from the multi-fidelity extension to the single-fidelity NMC. 
The optimal allocation $\mathcal{A}$ is found to be a generalized version of ACVIS (independent samples)~\cite{gorodetsky_generalized_2020} called the GISSR estimator~\cite{bomarito_optimization_2022}.
Table \ref{table:vars_reuse0} presents the MF-EIG estimator variance and variance reduction ratio relative to the single-fidelity NMC estimator,
averaged over 41 evenly-spaced designs across $\Design$. 
The MXMCPy tool takes just 1.38 seconds to find the optimal MF-EIG hyperparameters on a single 3.0 GHz Intel Xeon Gold 6154 processor.
Projected values are predicted by MXMCPy during hyperparameter optimization, and empirical values are obtained from repeating 50 estimation trials using fresh samples. 
As shown by the first two columns of Table \ref{table:vars_reuse0}, even with this na\"{i}ve-$\text{N}_{\text{in}}$ approach, 
6--8 times variance
reduction is achieved relative to single-fidelity NMC.

\begin{table}[htbp]
\centering
\begin{tabular}{ cccccc } 
 \hline 
 & \parbox[c]{20mm}{\strut\centering\textbf{Single-fidelity \\NMC}\strut} & 
 \parbox[c]{20mm}{\strut\centering\textbf{Na\"{i}ve}-$\mathrm{N}_{\text{in}}$ \\\textbf{(no reuse)}\strut} 
 & 
 \parbox[c]{20mm}{\strut\centering\textbf{Optimal}-$\mathrm{N}_{\text{in}}$ \\\textbf{(no reuse)}\strut} 
 & 
 \parbox[c]{20mm}{\strut\centering\textbf{Na\"{i}ve}-$\mathrm{N}_{\text{in}}$ \\\textbf{(reuse)}\strut} 
 & 
 \parbox[c]{20mm}{\strut\centering\textbf{Optimal}-$\mathrm{N}_{\text{in}}$ \\\textbf{(reuse)}\strut} 
 \\ 
 \hline \noalign{\smallskip}
 \parbox[c]{20mm}{\strut\centering Variance \\ (projected)\\[0.75em]} & $6.34\times 10^{-4}$ & $9.66\times 10^{-5}$ & $8.61\times 10^{-5}$ & $3.70\times 10^{-5}$ & $3.37\times 10^{-5}$ \\ 

 \parbox[c]{20mm}{\strut\centering Variance \\ (empirical)\\[0.75em]} & $7.11\times 10^{-4}$ & $9.22\times 10^{-5}$ & $8.59\times 10^{-5}$ & $3.61\times 10^{-5}$ & $3.18\times 10^{-5}$ \\ 

 \parbox[c]{20mm}{\strut\centering Variance \\ reduction ratio \\ (projected)\\[0.75em]} & -- & 6.56 & 7.36 & 17.16 & 18.30 \\ 

 \parbox[c]{20mm}{\strut\centering Variance \\ reduction ratio \\ (empirical)} & -- & 7.71 & 8.28 & 19.70 & 22.38 \\
 \noalign{\smallskip}\hline
\end{tabular}
\caption{Case 1. MF-EIG estimator variance and variance reduction ratio relative to the single-fidelity NMC estimator. Projected values are predicted from MXMCPy, and empirical values are obtained from repeating 50 estimation trials.}
\label{table:vars_reuse0}
\end{table}

In the second approach, the inner-loop sample sizes are optimized following \eqref{eqn:opt-nin}. 
The inner optimization, under a given $\mathrm{N}_{\text{in}}$, is solved using MXMCPy. 
The outer optimization, an integer program, is tackled using a grid search over reasonable values of $\mathrm{N}_{\text{in}}$; here we use a $50\times50$ grid over an integer-valued search space of $[25,4000]^{2}$. 
Grid search, however, is only practical when $M$ is relatively small and where the inner optimization is relatively fast. 
Advanced techniques of integer programming, and mixed-integer strategies that combine the inner and outer optimization problems, can be quite useful but are not explored in this paper.
The optimal allocation $\mathcal{A}$ is found to be a generalized version of MLMC~\cite{giles_multilevel_2015} called the WRDIFF estimator~\cite{bomarito_optimization_2022}. 

\begin{figure}[htbp]
    \centering
    \subfloat[\centering Computational cost (dashed) and correlation to $u_{0}$ (solid)]{{\includegraphics[width=0.485\textwidth]{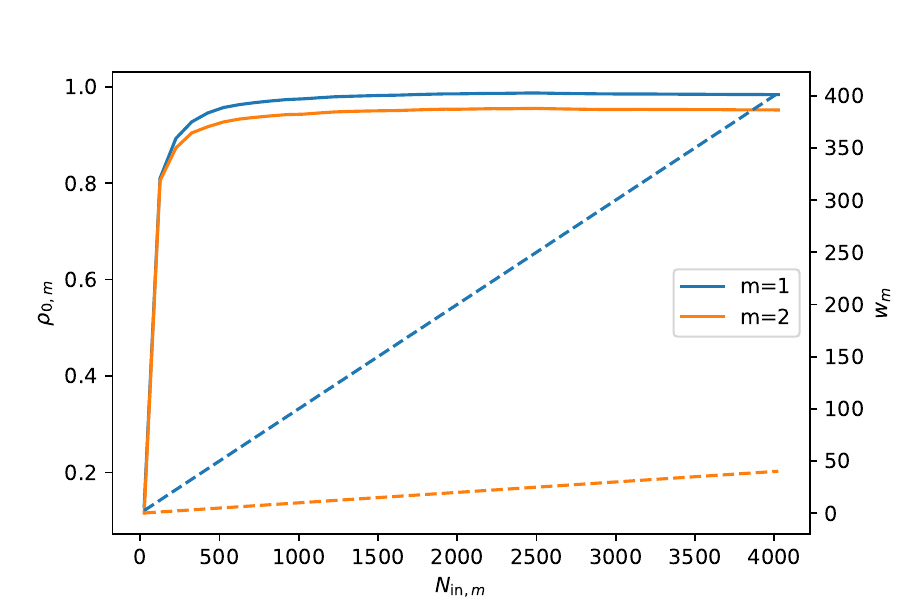} } \label{fig:costcorrs_reuse0_a}}%
    \subfloat[\centering Correlation between $u_{1}$ and $u_{2}$]{{\includegraphics[width=0.485\textwidth]{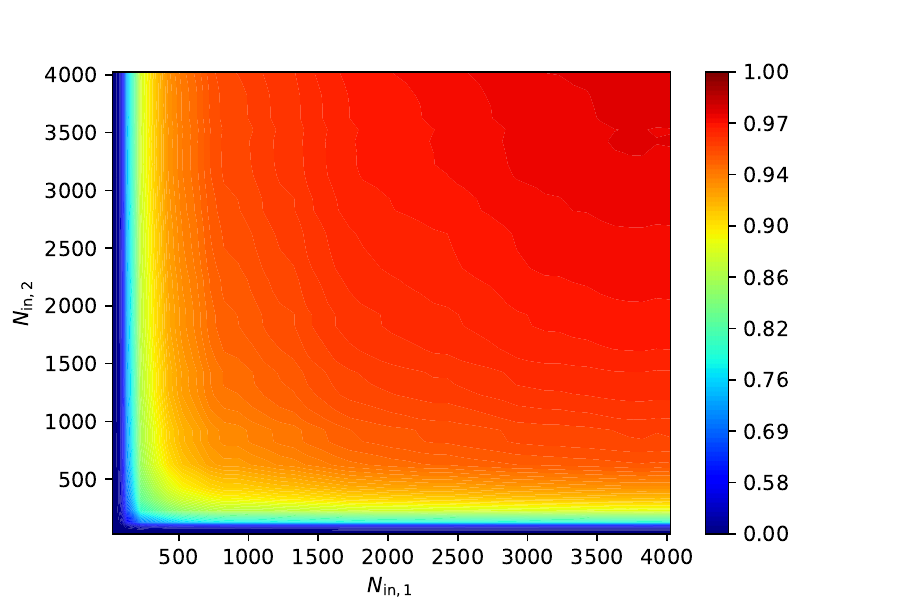} }\label{fig:costcorrs_reuse0_b}}%
    \caption{Case 1 (no reuse). Computational cost and design-averaged correlation as $N_{\text{in},1}$ and $N_{\text{in},2}$ change.
    (a) Low-fidelity log-ratio model computational costs (dashed lines) and correlations (solid lines) to the high-fidelity log-ratio model over inner-loop sample sizes (no sample reuse in the inner loop). The complex trade-off between correlations and costs dictate the best inner-loop sample sizes to use. 
    (b) Correlations between low-fidelity log-ratio model over inner-loop sample sizes (no sample reuse in the inner loop).}
    \label{fig:costcorrs_reuse0}%
\end{figure}

\begin{figure}[htbp]
    \centering
    \includegraphics[width=10cm]{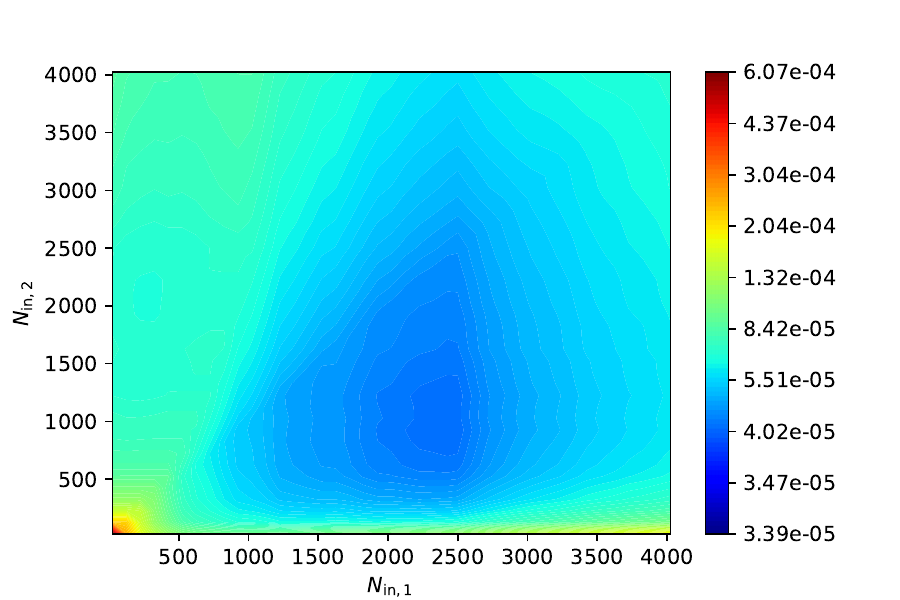}
    \caption{Case 1 (no reuse). Projected MF-EIG estimator variance across different possible low-fidelity inner-loop sample sizes,} $\mathrm{N}_{\text{in}}$. The contour levels use a power law normalization for enhanced visualization. 
    \label{fig:gridsearch_reuse0}
\end{figure}

The potential benefits from optimizing $\mathrm{N}_{\text{in}}$ can be seen in Figure \ref{fig:costcorrs_reuse0}. Figure~\ref{fig:costcorrs_reuse0_a} plots the computational cost of $u_1$ and $u_2$, and their design-averaged correlation to $u_0$, as $N_{\text{in},m}$ varies. It is evident that too small of a $N_{\text{in},m}$ results in low correlation, while excessively large $N_{\text{in},m}$ yields diminishing returns. 
Figure~\ref{fig:costcorrs_reuse0_b} further plots the design-averaged correlation between $u_1$ and $u_2$, revealing a similar trade-off between correlation and sample size. 
Figure \ref{fig:gridsearch_reuse0} shows the projected variance 
under different $\mathrm{N}_{\text{in}}$, where a region of low variance emerges around 
the optimum at $\mathrm{N}_{\text{in}} = [2500, 925]$. Very small values of $\mathrm{N}_{\text{in}}$, in particular, can lead to much higher variance than the optimal setting. 

\begin{figure}[htbp]
    \centering
    \includegraphics[width=10cm]{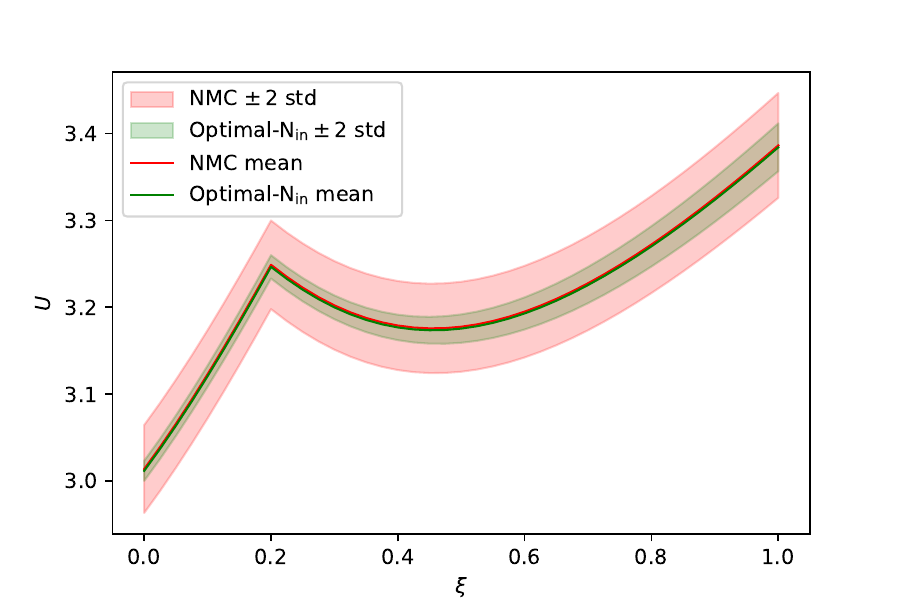}
    \caption{Case 1 (no reuse). Empirical estimator mean and $\pm 2$ standard deviation obtained from 50 repeated estimation trials using the single-fidelity NMC and optimal-$\mathrm{N}_{\text{in}}$ MF-EIG estimators.}
    \label{fig:acrossd_noreuse}
\end{figure}

Figure \ref{fig:acrossd_noreuse} shows the empirical estimator mean and $\pm 2$ standard deviation obtained from 50 repeated estimation trials using the single-fidelity NMC and the optimal-$\mathrm{N}_{\text{in}}$ MF-EIG estimators. 
In obtaining these results, we employ the common random number technique by fixing the random seed across $\Design$ when sampling $Z$ in the ACV outer loop---that is, the $z$ input on the left-hand side of \eqref{eqn:MF-EIG} is the same for every $\xi$ queried. This in turn leads to increased correlation among estimators across $\Design$ and in turn improving the smoothness of these curves.
Overall, the estimator standard deviation is much smaller for MF-EIG compared to single-fidelity NMC. 
Notably, the estimator standard deviation increases at larger values of $\design$, 
suggesting that MF-EIG may be further improved from $\design$-specific tuning. 
The excellent alignment in the mean curves confirms the unbiasedness of the MF-EIG estimator with respect to the high-fidelity mean. 
The design-averaged optimal-$\text{N}_{\text{in}}$ estimator variance and variance reduction ratio relative to single-fidelity NMC, both projected and empirical, are summarized in the third column of Table \ref{table:vars_reuse0}. Optimal-$\text{N}_{\text{in}}$ achieves around 7--9 times variance reduction relative to single-fidelity NMC, and a 6--12\% design-averaged improvement over na\"{i}ve-$\text{N}_{\text{in}}$.

\subsubsection{Sample reuse in the inner loop}\label{sec:reuse_toy}

We implement the sample-reuse strategy discussed in Section \ref{secn:reuse}, where  inner-loop samples of ${\Theta}$ are reused (rather than independently re-sampled each time) across utility models. 
The optimal allocations $\mathcal{A}$ for both na\"{i}ve-$\text{N}_{\text{in}}$ and optimal-$\text{N}_{\text{in}}$ estimators
are again the GISSR estimator~\cite{gorodetsky_generalized_2020}.
Overall, as shown by the last two columns of Table \ref{table:vars_reuse0}, sample reuse estimators achieve around 17--23 times variance reduction relative to single-fidelity NMC, and about double the reduction from their no-reuse counterparts.

\begin{figure}[htbp]
    \centering
    \subfloat[\centering Computational cost (dashed) and correlation to $u_{0}$ (solid)]{{\includegraphics[width=0.49\textwidth]{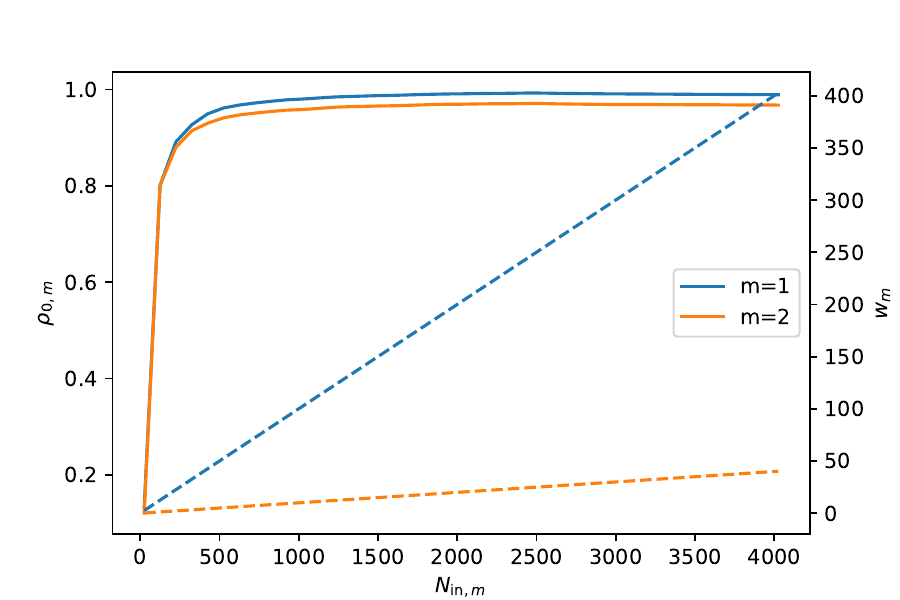} }\label{fig:corrcosts_reuse}}%
    \subfloat[\centering Correlation between $u_{1}$ and $u_{2}$]{{\includegraphics[width=0.49\textwidth]{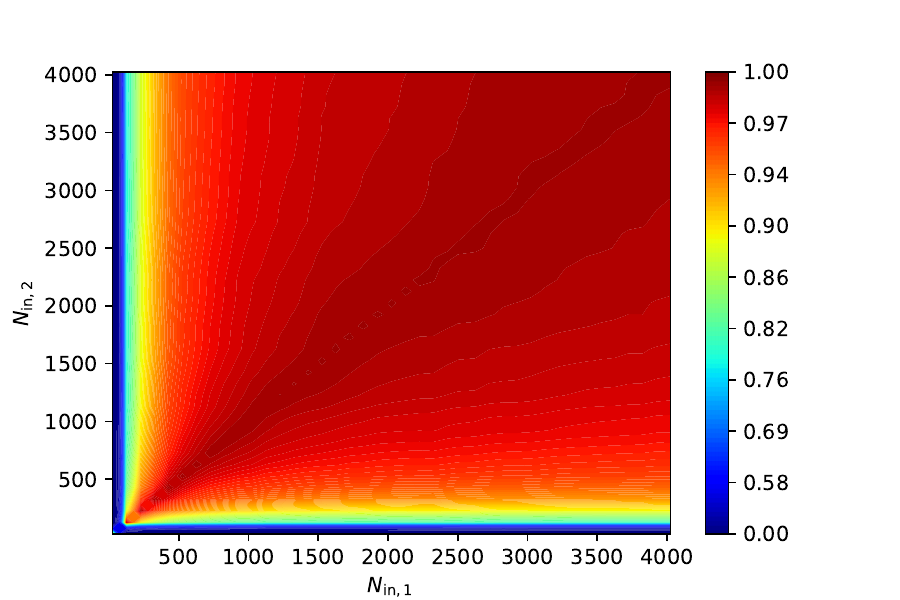} }\label{fig:corrs12_reuse}}%
    \caption{Case 1 (reuse). Computational cost and design-averaged correlation as $N_{\text{in},1}$ and $N_{\text{in},2}$ change. }%
    \label{fig:costcorrs_reuse}%
\end{figure}

\begin{figure}[htbp]
    \centering
    \includegraphics[width=10cm]{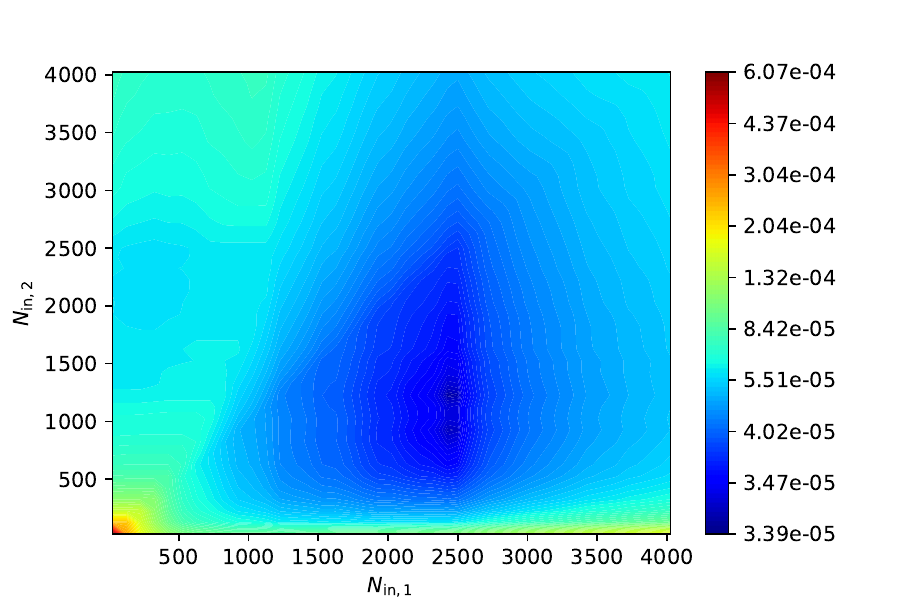}
    \caption{Case 1 (reuse). Projected MF-EIG estimator variance across different possible low-fidelity inner-loop sample sizes, $\mathrm{N}_{\text{in}}$. The contour levels use a power law normalization 
    for enhanced visualization.}
    \label{fig:gridsearch_reuse}
\end{figure}

\begin{figure}[htbp]
    \centering
    \includegraphics[width=10cm]{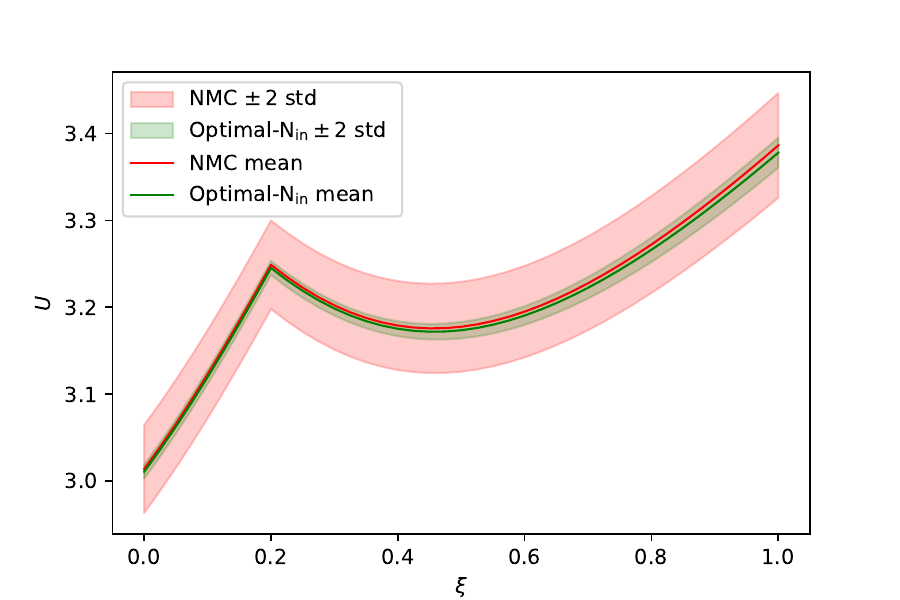}
    \caption{Case 1 (reuse). Empirical estimator mean and $\pm 2$ standard deviation obtained from 50 repeated estimation trials using the single-fidelity NMC and optimal-$\mathrm{N}_{\text{in}}$ MF-EIG estimators.}
    \label{fig:acrossd_reuse}
\end{figure}

Figure \ref{fig:costcorrs_reuse} plots the computational cost and design-averaged correlation trade-offs under reuse,
showing strengthened correlation compared to Figure \ref{fig:costcorrs_reuse0} especially between the $u_1$ and $u_2$ and at low values of $N_{\text{in},1}$ and $N_{\text{in},2}$.
Figure~\ref{fig:gridsearch_reuse} shows the projected variance under reuse. 
A low-variance region appears similar to that of Figure \ref{fig:gridsearch_reuse0}, but with lower $N_{\text{in},1}$ 
due to the bolstered correlation between $u_{1}$ and $u_{2}$ when $N_{\text{in},1}\approx N_{\text{in},2}$, as seen in Figure \ref{fig:corrs12_reuse}.
The optimum is at $\text{N}_{\text{in}}=[2425,1225]$, where $N_{\text{in},1}$ has decreased and $N_{\text{in},2}$ increased from the no-reuse case, indicating that more of the computational burden has been passed from $u_1$ to the less costly $u_2$ again due to their increased correlation. 
Figure \ref{fig:acrossd_reuse} shows the empirical estimator mean and $\pm 2$ standard deviation obtained from 50 repeated estimation trials using the single-fidelity NMC and optimal-$\mathrm{N}_{\text{in}}$ MF-EIG estimators under reuse. The estimator standard deviation for MF-EIG under reuse is further reduced from its no-reuse counterpart in Figure \ref{fig:acrossd_noreuse}.

\begin{figure}[htbp]
    \centering
    \includegraphics[width=10cm]{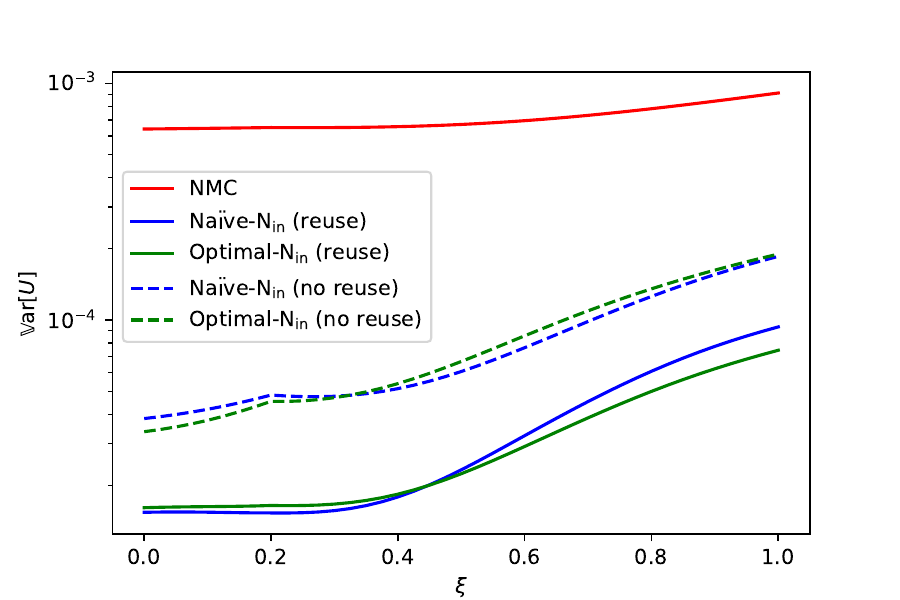}
    \caption{Case 1. Summary of estimator variance across $\Design$ for all estimators. }
    \label{fig:var_across_d}
\end{figure}

Figure \ref{fig:var_across_d} summarizes estimator variance across $\Design$ for all estimators, with and without reuse. It is evident that MF-EIG estimators perform much better than single-fidelity NMC, with sample reuse bringing notable improvements. 

\subsubsection{Scaled noise variant}
\label{sss:nonlinear_scaled}

\begin{figure}[htbp]
    \centering
    \subfloat[\centering Empirical estimator mean and $\pm 2$ standard deviation]{{\includegraphics[width=0.49\textwidth]{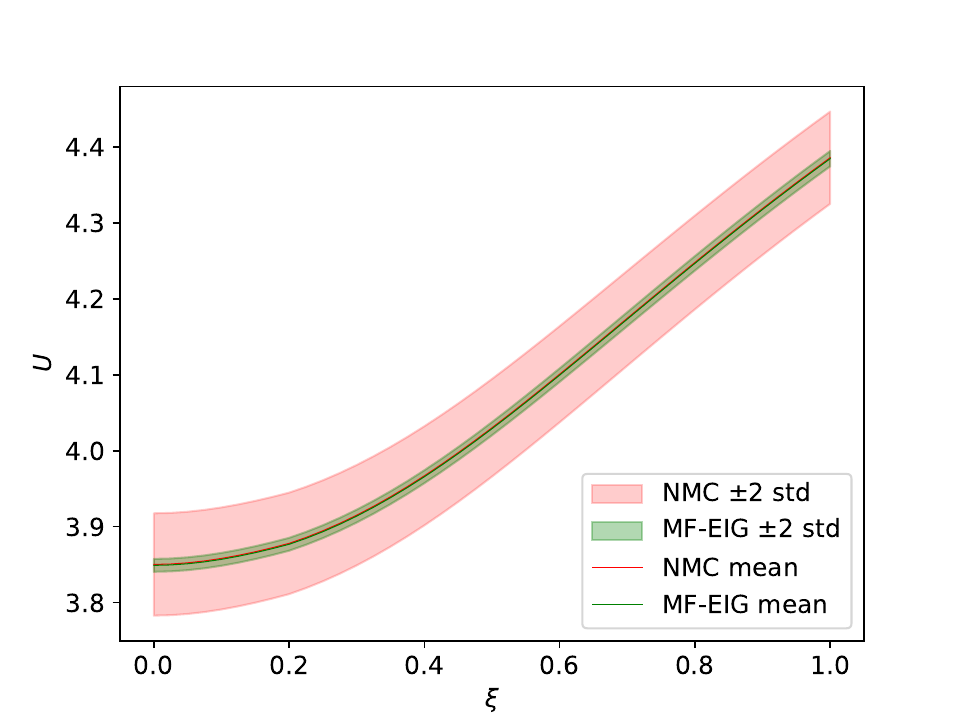} }\label{fig:u_d_mult}}%
    \subfloat[\centering Summary of estimator variance]{{\includegraphics[width=0.49\textwidth]{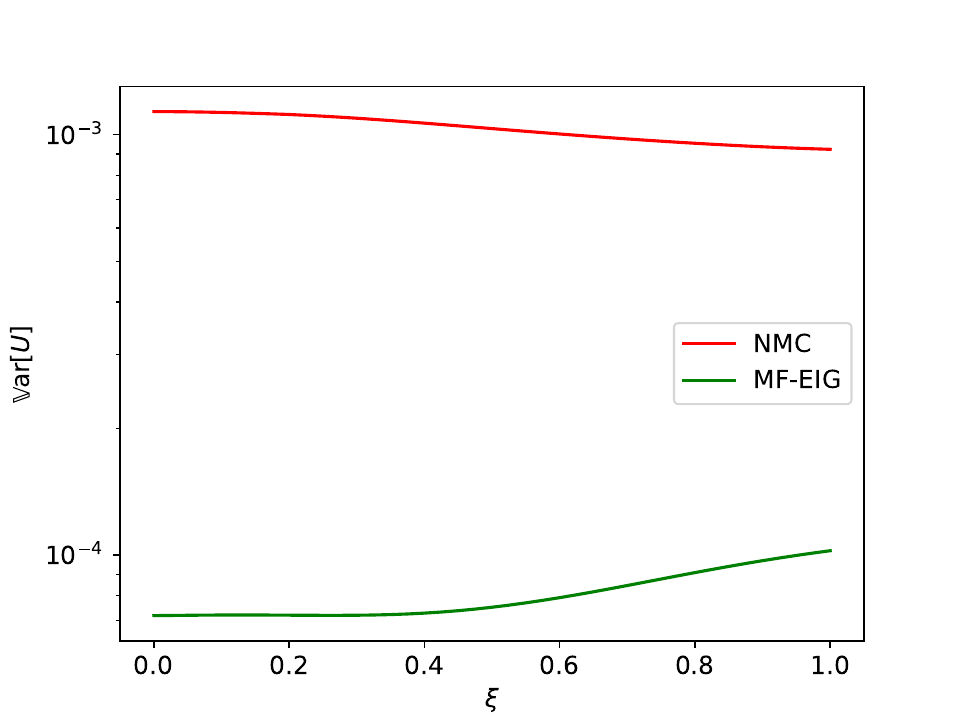} }\label{fig:var_across_d_mult}}%
    \caption{Case 1 (scaled noise). Empirical estimator results obtained from 50 repeated estimation trials using the single-fidelity NMC and na\"{i}ve-$\text{N}_{\text{in}}$ (with reuse) estimators. }%
    \label{fig:scaled_noise_plots}%
\end{figure}

We now illustrate MF-EIG on a scaled noise variant of Case 1, with all settings remaining the same except the data models are updated to:
\begin{align}
    Y_0 &= g_{0}(\Theta,\design) + g_{0}(\Theta,\design)\circ\Noise
    = \left(\Theta^{3}\design^{2} + \Theta \exp{( -|0.2-\design| )}\right) (1 +\Noise),\\
   Y_1 &= g_{1}(\Theta,\design) + g_{1}(\Theta,\design)\circ\Noise = \left(k_{1}\,\Theta^{2.5}\design^{1.75} + \Theta \exp{( -|0.2-\design| )} \right)(1+ \Noise), \\
   Y_2 &= g_{2}(\Theta,\design) + g_{2}(\Theta,\design)\circ\Noise = \left(k_{2}\,\Theta^{2}\design^{1.5} + \Theta \exp{( -|0.2-\design| )}\right) (1+\Noise).
\end{align}
The optimal allocation $\mathcal{A}$ is again the GISSR estimator \cite{bomarito_optimization_2022}.
Figure \ref{fig:u_d_mult} shows the empirical estimator mean and $\pm 2$ standard deviation obtained from 50 repeated estimation trials using the single-fidelity NMC and na\"{i}ve-$\text{N}_{\text{in}}$ (with reuse) estimators. 
Remarkable, with a seemingly minor alteration to the noise structure, 
the EIG trend differs substantially from the additive noise case in Figure~\ref{fig:acrossd_noreuse}, as the local maximum at $\design=0.2$ is now completely absent.
Additionally, the identical mean curves strikingly exemplifies the unbiasedness of the MF-EIG estimator.
The variance reduction achieved by MF-EIG remains similar, as summarized in
Figure \ref{fig:var_across_d_mult}.
The empirical design-averaged variance for the na\"{i}ve-$\text{N}_{\text{in}}$ MF-EIG estimator is
$8.04 \times 10^{-5}$, compared to 
$1.03 \times 10^{-3}$ 
for the
single-fidelity NMC estimator, equating to a 12.86 times variance reduction.

\subsection{Case 2: turbulent flow problem}

Modeling turbulent fluid flows is an important and challenging problem in computational aerospace and mechanical engineering. 
The underlying flow physics are governed by the Navier--Stokes partial differential equations (PDEs). Solving these PDEs at scale-resolving resolutions, especially for three-dimensional, high-Reynolds-number flows, typically requires an extremely large number of grid points (e.g., $10^{10}$), making such computations prohibitively expensive.

Alternatively, Reynolds-averaged Navier--Stokes (RANS) models have gained popularity due to their computational efficiency and ability to achieve reasonable accuracy. Rather than directly resolving the time-fluctuating components of turbulence, RANS employs eddy-viscosity-based closure models that approximate turbulence as a function of mean-flow quantities. 
These turbulence closure models are predominantly empirical, with a set of unknown parameters that must be estimated or inferred using experimental data \cite{xiao_quantification_2019}. 
Traditionally, these parameters are determined through trial-and-error, heuristics, or calibrating against
fundamental flow experiments (e.g., simple shear flows) \cite{shirzadi_rans_2020}.

It is often the case that RANS turbulence closure models need to be re-calibrated using relevant experimental data to better align with specific use cases.
Given that high-Reynolds-number experiments are typically costly and challenging to conduct, optimizing their design through OED becomes crucial. Moreover, while RANS is computationally efficient compared to higher-fidelity approaches, solvers with sufficiently fine grid resolutions can still be computationally expensive, making multi-fidelity OED particularly appealing for this context. 

\subsubsection{SST turbulence closure model}

We focus on a two-dimensional flow over a zero-gradient flat plate, a common verification case. The computational domain is $[-0.5,2]\times[0,1]$, with inlet boundary condition on the left edge of the domain, outlet boundary conditions on the top and right edges of the domain, and adiabatic no-slip wall boundary condition at the bottom of the domain where the plate is located.
The flow has Reynolds number $5\times10^6$ and is modeled using steady RANS with the two-equation shear stress transport (SST) turbulence closure model~\cite{menter_two-equation_1994}.
This closure model blends a near-wall $k$-$\omega$ formulation with a more conventional $k$-$\epsilon$ formulation in the free-stream regime.
The full closure model can be expressed using tensor notation as:
\begin{align}\label{eqn:sst}
    \frac{\partial (\rho k)}{\partial t} + \frac{\partial (\rho u_{j} k)}{\partial x_{j}} &= P_{1} - \beta^{\ast}\rho\omega k + \frac{\partial}{\partial x_{j}}\left[ (\mu + \sigma_{k}\mu_{t}) \frac{\partial k}{\partial x_{j}} \right],\\[6pt]
    \frac{\partial(\rho\omega)}{\partial t} + \frac{(\partial \rho u_{j} \omega) }{\partial x_{j}} &= \frac{\rho \gamma}{\mu_{t}} P_{1} - \beta\rho\omega^{2} + \frac{\partial}{\partial x_{j}} \left[ (\mu + \sigma_{\omega}\mu_{t})\frac{\partial\omega}{\partial x_{j}} \right] + 2(1-F_{1}) \frac{\rho \sigma_{\omega 2}}{\omega} \frac{\partial k}{\partial x_{j}} \frac{\partial \omega}{\partial x_{j}},
\end{align}
where $\rho$ is the density, $u$ is the velocity, $k$ is the kinetic energy, $\omega$ is the specific rate of dissipation of $k$ into thermal energy, $P_1$ is the production term, $\mu$ is the molecular viscosity, $\mu_t$ is the turbulent eddy viscosity, and $\beta^\ast$, $\beta$, $\sigma_k$, $\sigma_{\omega2}$, and $\gamma$ are tunable parameters.
$P_1$ is defined as
\begin{align}
    P_{1} = \tau_{ij} \frac{\partial u_{i}}{\partial x_{j}} \hspace{3em} &\text{with} \quad
    \tau_{ij} = \mu_{t}\left(2S_{ij} - \frac{2}{3}\frac{\partial u_{k}}{\partial x_{k}} \delta_{ij}\right) - \frac{2}{3}\rho k \delta_{ij} \\ &\text{and}\quad
    S_{ij} = \frac{1}{2}\left(\frac{\partial u_{i}}{\partial x_{j}} + \frac{\partial u_{j}}{\partial x_{i}}\right), \nonumber
\end{align}
where $\tau_{ij}$ is the Reynolds shear stress tensor, $\delta_{ij}$ is the Kronecker delta function, and $S_{ij}$ is the strain rate tensor.
$\mu_t$ is defined as
\begin{align}
    \mu_{t} = \frac{\rho a_{1} k}{\max (a_{1}\omega, \Omega F_{2})},
\end{align}
where $\Omega$ is the vorticity magnitude, $a_1$ is another tunable parameter, and $F_2$ is given by
\begin{align}
    F_2 = \text{tanh}(\Pi^2) \hspace{3em} \text{with} \quad \Pi=\max(2\Gamma_3,\Gamma_1),\quad \Gamma_{1} = \frac{500\mu}{d^{2}\omega\rho}, \quad \Gamma_{3} = \frac{\sqrt{k}}{\beta^{*}\omega d},
\end{align}
with $d$ being the distance to the closest surface.

Furthermore, a subset of parameters, $\phi := \{\beta, \sigma_{k}, \sigma_{\omega}, \gamma\}$, are the result of blending $\phi_{1} := \{\beta_{1}, \sigma_{k1}, \sigma_{\omega1}, \gamma_{1}\}$ and $\phi_{2}:=\{\beta_{2}, \sigma_{k2}, \sigma_{\omega2}, \gamma_{2}\}$ originating from the original $k$-$\omega$ and $k$-$\epsilon$ models, respectively. The blending is done such that the near-wall dynamics are dictated by $\phi_{1}$ and the free-stream dynamics are dictated by $\phi_{2}$, controlled by a blending function $F_1$:
\begin{align}\label{eqn:blend}
    \phi = F_{1}\phi_{1} + (1-F_{1})\phi_{2},
\end{align}
where
\begin{align}
    F_{1} = \tanh{(\Gamma^{4})} \hspace{3em} \text{with} \quad &\Gamma = \min \left( \max (\Gamma_{1},\Gamma_{3}), \Gamma_{2}\right),
    \\
    &\Gamma_{2} = \frac{4\rho\sigma_{\omega2}k}{d^{2}(\text{CD}_{k\text{-}\omega})},\nonumber
    \\[2pt]  
    &\text{CD}_{k\text{-}\omega} = \max \left(  \frac{2\rho\sigma_{\omega2}}{\omega} \frac{\partial k}{\partial x_{i}} \frac{\partial \omega}{\partial x_{i}}, 10^{-20}\right).\nonumber
\end{align}
The scaling coefficients $\gamma_{1}$ and $\gamma_{2}$ are themselves functions of the other parameters:
\begin{align}
    \gamma_{1} = \frac{\beta_{1}}{\beta^{*}} - \sigma_{\omega1} \frac{k^{2}}{\sqrt{\beta^{*}}}, \quad \quad
    \gamma_{2} = \frac{\beta_{2}}{\beta^{*}} - \sigma_{\omega2} \frac{k^{2}}{\sqrt{\beta^{*}}}.
\end{align}

In the end, there are only 8 degrees of freedom for the tunable parameters of the SST turbulence closure model:
$a_{1}, \beta^{\ast}, \beta_{1}, \beta_{2}, \sigma_{k1}, \sigma_{k2}, \sigma_{\omega1}, \sigma_{\omega2}$. 

We have at our disposal four numerical flow solvers, each employing a computational grid of different resolution (see Table \ref{table:forward-models})
adopted from \cite{rumsey2018turbulence}. 
The grids are also nested, i.e., the nodes from the coarser grids are always contained in the finer grids.
The solvers are implemented using the Stanford University Unstructured (SU2) framework
\cite{economon_su2_2016} employing a second-order, upwind convective scheme and implicit Euler time integration. 
The computational cost of each solver (see Table \ref{table:forward-models}) is determined by measuring its approximate time to acceptable convergence for a single flow solve, performed on the University of Michigan Great Lakes HPC Cluster using a single core of a 3.0 GHz Intel Xeon Gold 6154 processor. 

\begin{table}[htbp]
\centering
\begin{tabular}{ crrr} 
 \hline
 \textbf{Flow solver} & \textbf{Grid} & \textbf{Total grid nodes} & \textbf{Computational cost (seconds)} \\ 
 \hline \noalign{\smallskip}
 S$_{0}$  & $273 \times 193$ & 52,689 & 1778.5 \\ 

 S$_{1}$  & $137 \times 97$ & 13,289 & 298.2 \\ 

 S$_{2}$  & $69 \times 49$ & 3,381 &  27.4 \\ 

 S$_{3}$  & $35 \times 25$ & 875 &  1.9 \\
 \noalign{\smallskip} \hline
\end{tabular}
\caption{Case 2. Flow solver details. The computational cost of each solver is the $w^g_m$ in the MF-EIG framework.}
\label{table:forward-models}
\end{table}

\subsubsection{Optimal experimental design}

We seek to design the location of a pressure probe whose measurement maximizes the EIG in the SST closure model parameters.
The high-fidelity data model for this problem is:
\begin{align}
    Y_0 = g_{0}(\Theta,\design) + \Noise,
\end{align}
where $\Theta = [ a_{1}, \beta^{\ast}, \beta_{1}, \beta_{2}, \sigma_{k1}, \sigma_{k2}, \sigma_{\omega1}, \sigma_{\omega2} ]$ is the vector of SST closure model parameters, $\design$ is the two-dimensional design location for placing a pressure probe, $g_0$ is the forward operator that predicts the mean-flow pressure magnitude at the probe location using the flow solution obtained from solver S$_0$, 
$Y_0$ is the measured mean-flow pressure magnitude, and $\Noise \sim \mathcal{N}(0,(10^{-3})^2)$ is a Gaussian data noise.
The parameter priors are composed of independent uniform distributions, with upper and lower bounds adopted from \cite{zhang_uncertainty_2022} and listed in Table~\ref{tab:priors}.
We restrict the design space $\Design$ to discrete locations of the $875$ nodes
on the coarsest grid (i.e., of S$_3$).
Since the grids are nested, these locations are directly accessible in all solvers without needing interpolation. 
The low-fidelity data models have the same structure:
\begin{align}
    Y_m = g_{m}(\Theta,\design) + \Noise,\quad m=1,2,3,
\end{align}
where $g_m$ is the low-fidelity forward operator that predicts the mean-flow pressure magnitude at the probe location using the flow solution obtained from solver S$_m$. The cost of evaluation each forward operator $g_m$ is the same as the solver computational cost listed in the last column of Table~\ref{table:forward-models}.

We note that in this problem setup, a single flow solve produces pressure predictions simultaneously at every single design location. As such, the OED problem can be solved via a simple grid search through $\Design$ without the additional computational costs of separate EIG estimates at each design.

\begin{table}[htbp]
    \centering
    \begin{tabular}{c c c}
    \hline
        \textbf{Parameter} & \textbf{Lower bound} & \textbf{Upper bound} \\
    \hline \noalign{\smallskip}
        $a_{1}$ & 0.3 & 0.4 \\
        $\beta^{\ast}$ & 0.0784 & 0.1024 \\
        $\beta_{1}$ & 0.06 & 0.09 \\
        $\beta_{2}$ & 0.07 & 0.1 \\
        $\sigma_{k1}$ & 0.7 & 1.0 \\
        $\sigma_{k2}$ & 0.8 & 1.2 \\
        $\sigma_{\omega1}$ & 0.3 & 0.7 \\
        $\sigma_{\omega2}$ & 0.7 & 1.0 \\
    \noalign{\smallskip} \hline
    \end{tabular}
    \caption{Case 2. Lower and upper bounds of independent uniform prior distributions for the SST closure model parameters.}
    \label{tab:priors}
\end{table}

\subsubsection{EIG estimation}\label{sec:rans-mfeig}

We employ three EIG estimators for this case:  single-fidelity NMC, na\"{i}ve-$\mathrm{N}_{\text{in}}$ MF-EIG, and optimal-$\mathrm{N}_{\text{in}}$ MF-EIG. 
Based on the findings from Section \ref{sec:reuse_toy}, we implement all MF-EIG estimators with inner-loop sample reuse across their utility models. 
We also employ the common random number technique by fixing the random seed across $\Design$ when sampling $Z$ in the ACV outer loop.
All MF-EIG estimators are then formed from a total computational budget of $w_{\text{budget}} = 1.79 \times 10^{7}$ seconds.
Furthermore, we fix the inner-loop sample size of the high-fidelity utility model, $u_0$, to $N_{\text{in},0}=136$. This choice balances bias control with our total computational budget, while enabling a reasonable outer-loop sample size if using standard NMC.
Under the same budget and inner-loop configuration, the single-fidelity NMC baseline allows for an outer-loop sample size of $N_{\text{out}} = 74$ (after rounding up).

\begin{figure}[htbp]
    \centering
    \subfloat[\centering Utility model evaluations]{{\includegraphics[width=0.47\textwidth]{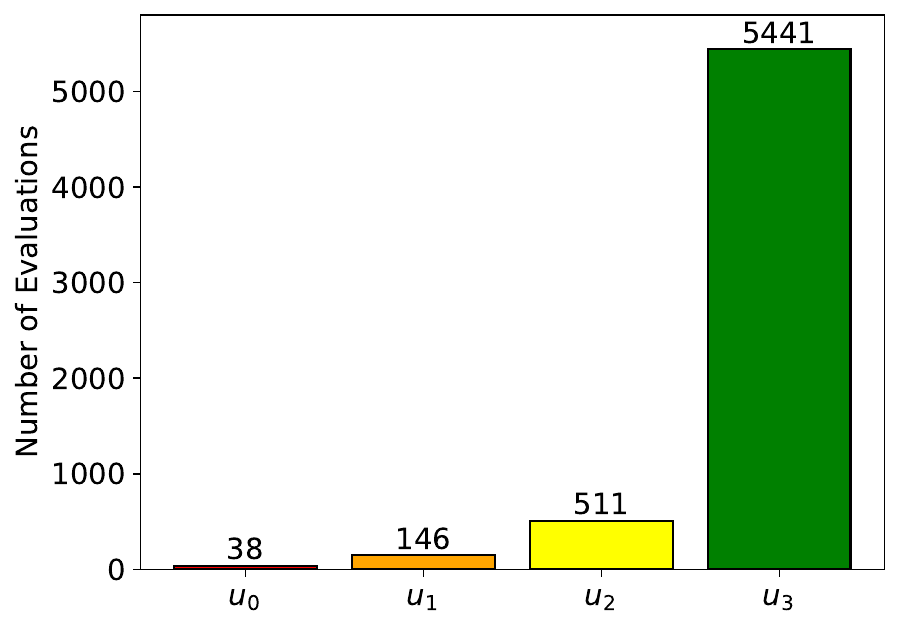} }\label{fig:naive_evals_hist}}
    \hspace{1em}
    \subfloat[\centering Utility model total computational cost]{{\includegraphics[width=0.47\textwidth]{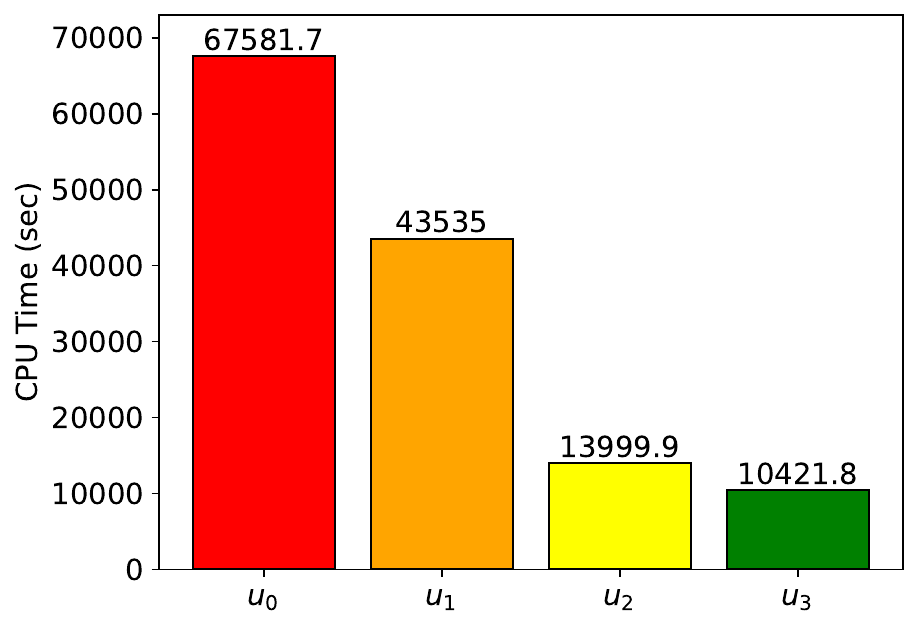} }\label{fig:naive_cost_hist}}
    \caption{Case 2. Total number of evaluations and total computational cost of utility models for the na\"{i}ve-$\mathrm{N}_{\text{in}}$ MF-EIG estimator under its optimal allocation $\mathcal{A}$.}%
    \label{fig:naive_cost_evals}%
\end{figure}

The na\"{i}ve-$\mathrm{N}_{\text{in}}$ MF-EIG estimator fixes $N_{\text{in},m} = 136$ for all $m$. 
We employ $N_{\text{pilot}}=136$ pilot samples to estimate the sample covariance following~\eqref{eqn:cov} at each of the $N_{\design}=875$ designs in $\Design$, and then average them to obtain the design-averaged utility model covariance in \eqref{eqn:avg-cov}.
Since each flow solve produces $g_m$ for all designs simultaneously through its computational grid, the pilot phase entails only $N_{\text{pilot}}$ flow solves of each solver rather than $N_{\text{pilot}}\times N_{\design}$. 
The computational cost of each utility model is estimated from the average run time over these pilot simulations. 
Correlation and computational cost information for the utility models are shown in Table~\ref{table:logratio_models_naive}.
From solving \eqref{eqn:mxmcpy}--\eqref{eqn:mxmcpy_constraint}, the optimal allocation $\mathcal{A}$ 
is found to be a generalized version of the MFMC estimator~\cite{peherstorfer_optimal_2016} called the ACVMF estimator~\cite{gorodetsky_generalized_2020}. 
The MXMCPy tool takes 10.89 seconds to find the optimal MF-EIG hyperparameters using a single 3.0 GHz Intel Xeon Gold 6154 processor, a negligible cost in comparison to the costs of utility model evaluations.
The resulting total number of evaluations and total computational cost of utility models under this allocation are depicted in Figure \ref{fig:naive_cost_evals}.
We see the allocation favors many more lower-fidelity utility model evaluations, although the total cost remains higher for the higher-fidelity models.

\begin{table}[htbp]
\centering
\begin{tabular}{ c c c c } 
 \hline
 \multicolumn{1}{p{2.5cm}}{\centering \textbf{Utility \\ model} }
 &  \multicolumn{1}{p{2.5cm}}{\centering \textbf{Underlying \\ solver} }
 & 
 \multicolumn{1}{p{3cm}}{\centering \textbf{Correlation \\ to} ${u_{0}}$ }
 & \multicolumn{1}{p{3cm}}{\centering \textbf{Comp. cost} \\ $w_m$ \textbf{(seconds)}} \\ 
 \hline \noalign{\smallskip}
 $u_{0}$  & S$_{0}$ & 1 & 
 $2.42\times10^{6}$ \\ 

 $u_{1}$  & S$_{1}$ & 0.995 & 
 $4.06\times10^{4}$ \\ 

 $u_{2}$  & S$_{2}$ & 0.981 &  
 $3.73\times10^{3}$ \\ 

 $u_{3}$  & S$_{3}$ & 0.953 &  
 $2.60\times10^{3}$ \\
 \noalign{\smallskip} \hline
\end{tabular}
\caption{Case 2. Correlation and computational cost of utility models for the na\"{i}ve-$\mathrm{N}_{\text{in}}$ MF-EIG estimator.}
\label{table:logratio_models_naive}
\end{table}

\begin{figure}[htbp]
    \centering
    \subfloat[\centering Utility model evaluations]{{\includegraphics[width=0.47\textwidth]{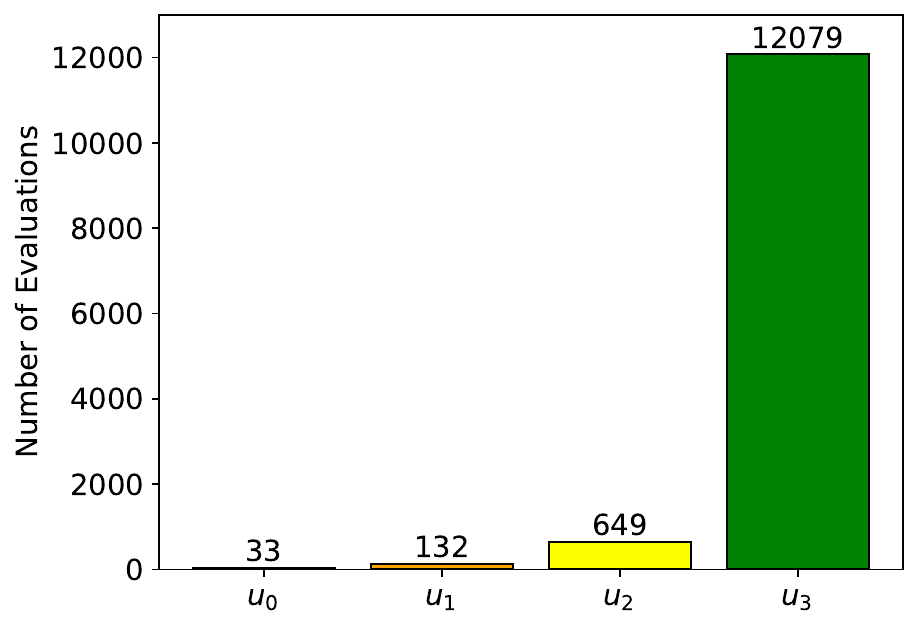} }\label{fig:opt_evals_hist}}
    \hspace{1em}
    \subfloat[\centering Utility model total computational cost]{{\includegraphics[width=0.47\textwidth]{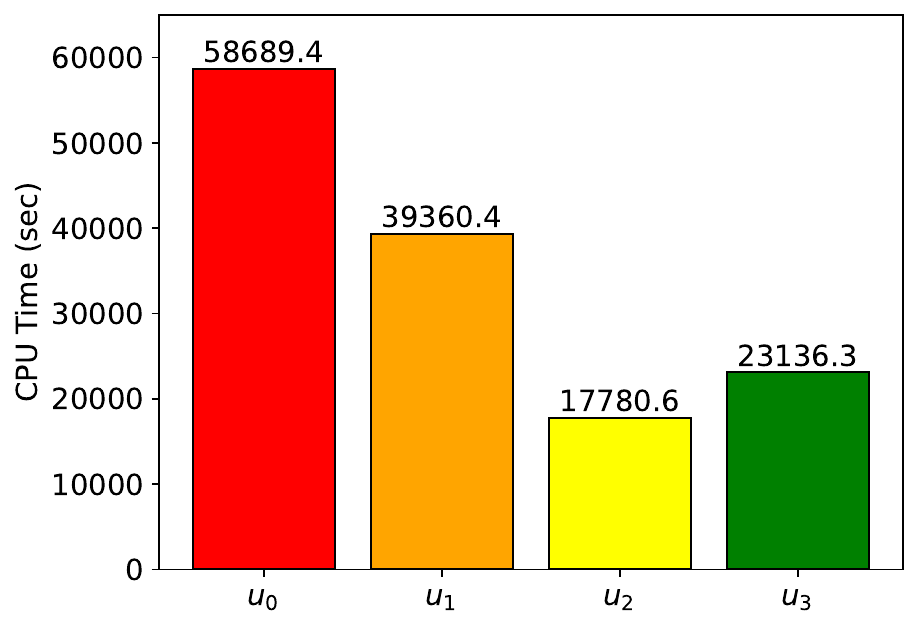} }\label{fig:opt_cost_hist}}
    \caption{Case 2. Total number of evaluations and total computational cost of utility models for the optimal-$\mathrm{N}_{\text{in}}$ MF-EIG estimator under its optimal allocation $\mathcal{A}$.}%
    \label{fig:opt_cost_evals}%
\end{figure}

The optimal-$\mathrm{N}_{\text{in}}$ MF-EIG estimator optimizes $\mathrm{N}_{\text{in}}$ over an integer-valued search space of $[5, 136]^3$ following \eqref{eqn:opt-nin}.
The pilot sampling process is identical to the na\"{i}ve-$\mathrm{N}_{\text{in}}$ case.
To accelerate the computation, we relax the optimization problem to real numbers in $[5, 136]^3$, and solve it using the Divided Rectangles (DIRECT) \cite{Gablonsky2001} global optimization implementation from the SciPy \cite{2020SciPy-NMeth} module. The result is then rounded to the nearest feasible integer. 
Correlation and computational cost information for the utility models are shown in Table~\ref{table:logratio_models_opt}.
From solving \eqref{eqn:mxmcpy}--\eqref{eqn:mxmcpy_constraint}, the optimal allocation $\mathcal{A}$ 
is found to be a generalized version of the ACVIS estimator (an even more generalized version than GISSR) called the GISMR estimator \cite{bomarito_optimization_2022}. 
The resulting total number of evaluations and total computational cost of utility models under this allocation are depicted in Figure \ref{fig:opt_cost_evals}.
Compared to Figure \ref{fig:naive_cost_evals}, even more model evaluations and computational cost are now shifted from higher- to lower-fidelity utility models, taking advantage of their correlation.
\begin{table}[htbp]
\centering
\begin{tabular}{ c c c c c } 
 \hline
 \multicolumn{1}{p{2cm}}{\centering \textbf{Utility \\ model} }
 &  \multicolumn{1}{p{2.5cm}}{\centering \textbf{Underlying \\ solver} }
 & 
 \multicolumn{1}{p{2.5cm}}{\centering \textbf{Correlation \\ to} ${u_{0}}$ }
 & \multicolumn{1}{p{2.5cm}}{\centering \textbf{Optimal} \\ $N_{\text{in},m}$} 
 & \multicolumn{1}{p{2.5cm}}{\centering \textbf{Comp. cost} \\ $w_m$ \textbf{(seconds)}} \\ 
 \hline \noalign{\smallskip}
 $u_{0}$  & S$_{0}$ & 1 & 136 & 
 $2.42\times10^{6}$ \\ 

 $u_{1}$  & S$_{1}$ & 0.993 & 115 & 
 $3.56\times10^{4}$ \\ 

 $u_{2}$  & S$_{2}$ & 0.978 & 114 & 
 $3.19\times10^{3}$ \\ 

 $u_{3}$  & S$_{3}$ & 0.881 & 29 & 
 $5.02\times10^{1}$ \\
 \noalign{\smallskip} \hline
\end{tabular}
\caption{Case 2. Correlation and computational cost of utility models for the optimal-$\mathrm{N}_{\text{in}}$ MF-EIG estimator.}
\label{table:logratio_models_opt}
\end{table}

\subsubsection{Results and discussion}

Table \ref{table:RANS_results} presents the MF-EIG estimator variance and variance reduction ratio relative to the single-fidelity NMC, averaged over all 875 designs across $\Design$ and at the optimal design $\design^{\ast}$. Projected values are predicted by MXMCPy during hyperparameter optimization, and empirical values are obtained from repeating 22 estimation trials using fresh samples. 
Both the na\"{i}ve-$\mathrm{N}_{\text{in}}$ and optimal-$\mathrm{N}_{\text{in}}$ MF-EIG estimators improve substantially from the single-fidelity NMC, achieving more than one-order-of-magnitude (19--71 times) variance reduction. 
Empirical variance reduction tends to be even greater than projected by MXMCPy. 
Importantly, the benefit received at the optimal design, $\design^{\ast}$, is much higher than those averaged across all designs, achieving roughly two-orders-of-magnitude (94--233 times) variance reduction. 
Similar to Case 1 in Section \ref{sec:toy}, optimal-$\mathrm{N}_{\text{in}}$ MF-EIG sees a modest 14--21\% design-averaged improvement over na\"{i}ve-$\mathrm{N}_{\text{in}}$, and a 151\% improvement at $\design^{\ast}$. 

\begin{table}[htbp]
\centering
\begin{tabular}{ cccc } 
 \hline 
 & \multicolumn{1}{p{3cm}}{\centering \textbf{\textbf{Single-fidelity\\ NMC}} }
 & \textbf{Na\"{i}ve}-$\mathrm{N}_{\text{in}}$
 & \textbf{Optimal}-$\mathrm{N}_{\text{in}}$  \\ 
 \hline \noalign{\smallskip}
 \parbox[c]{65mm}{\strut Variance (projected)\strut} & $3.23 \times 10^{-3}$ & $1.67 \times 10^{-4}$ & $1.40 \times 10^{-4}$  \\ 

 \parbox[c]{65mm}{\strut Variance (empirical)\strut} & $9.24 \times 10^{-3}$ & $1.50 \times 10^{-4}$ & $1.32 \times 10^{-4}$  \\ 

 \parbox[c]{65mm}{\strut Variance at $\design^{\ast}$ (empirical)} & $2.43 \times 10^{-2}$ & $2.58 \times 10^{-4}$ & $1.04 \times 10^{-4}$  \\ 

 \parbox[c]{65mm}{\strut Variance reduction ratio (projected)\strut} & -- & 19.19 & 23.07  \\ 

 \parbox[c]{65mm}{\strut Variance reduction ratio (empirical)\strut} & -- & 61.48 & 70.19  \\

 \parbox[c]{65mm}{\strut Variance reduction ratio at $\design^{\ast}$ (empirical)\strut} & -- & 94.24 & 233.12 \\
 \noalign{\smallskip}\hline
\end{tabular}
\caption{Case 2. MF-EIG variance and variance reduction ratio relative to the single-fidelity NMC estimator. Projected values are predicted by MXMCPy, and empirical values are obtained from repeating 22 estimation trials.}
\label{table:RANS_results}
\end{table}

\begin{figure}[htbp]
    \centering
    \subfloat[\centering Optimal-$\mathrm{N}_{\text{in}}$ MF-EIG]{{\includegraphics[width=0.49\textwidth]{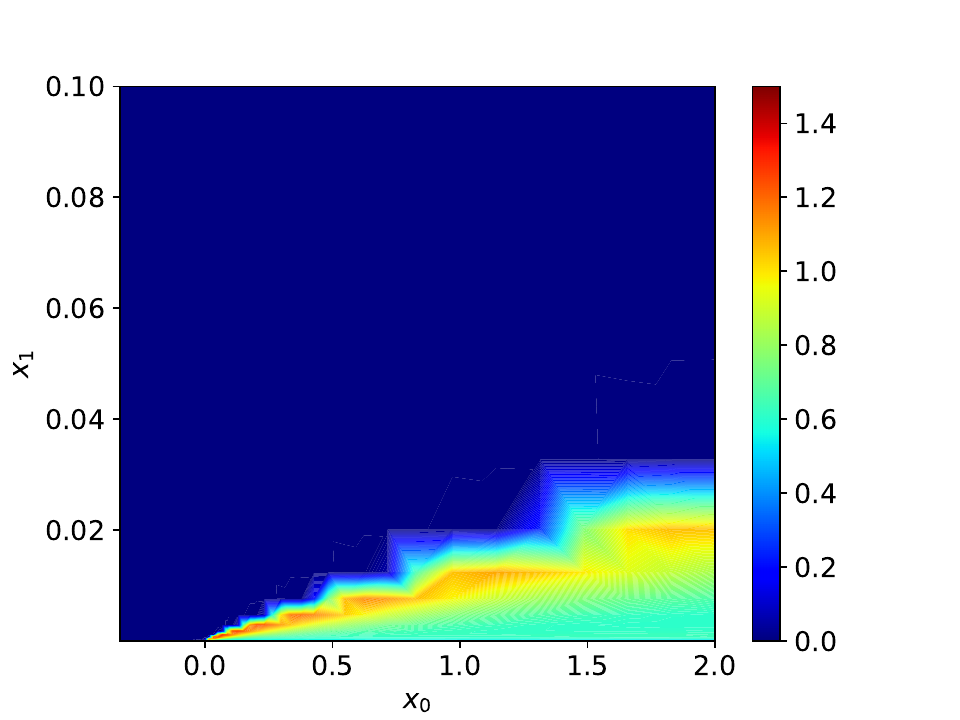} }\label{fig:contoursa}}
    \subfloat[\centering Optimal-$\mathrm{N}_{\text{in}}$ MF-EIG in semi-log $x_1$-axis]{{\includegraphics[width=0.49\textwidth]{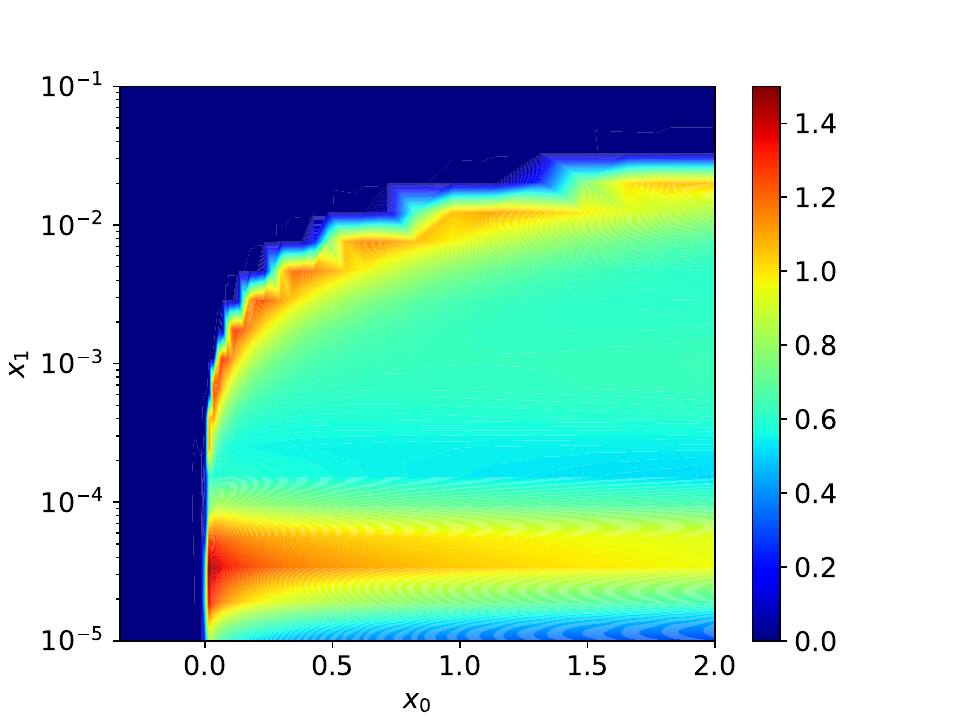} }\label{fig:contoursb}}
    \caption{Case 2. Empirical estimator mean obtained from 22 repeated estimation trials using optimal-$\mathrm{N}_{\text{in}}$ MF-EIG estimator. 
    The flow is from left to right, with an adiabatic flat plate starting at $x_{0}=0$ on the bottom of the domain triggering a turbulent boundary layer. The full computational domain extends to $x_{1}=1$, and these plots focus on $x_{1}\in[0,0.1]$ because regions further away all have zero EIG.}%
    \label{fig:contours}%
\end{figure}

Both the MF-EIG estimate itself 
and the estimator variance reduction can vary drastically across $\Design$. 
Figure~\ref{fig:contoursa} shows the empirical estimator mean obtained from 22 repeated estimation trials using the optimal-$\mathrm{N}_{\text{in}}$ MF-EIG estimator across a zoomed-in portion of $\Design$ close to the flat plate; Figure \ref{fig:contoursb} further plots it in semi-log $x_1$-axis (vertical axis) to accentuate the pattern closer to the core region of turbulent flow.
Recall that the full computational domain extends to $x_1=1$, and these plots focus on $x_1\in[0,0.1]$ because regions further away all have zero EIG.
The reason for zero EIG is that regions far away from the plate are not affected by, and thus do not carry information about, the turbulence phenomenon. 
The most informative design locations lie in the region just above the leading edge of the flat plate and near the apparent boundary layer. 
One possible explanation is that this region is greatly affected
by the SST closure model that controls the relative balance between free-stream and near-wall turbulence formulations.
The optimal design location, interestingly,
resides inside the boundary layer rather than on or near the boundary layer edge. 

\begin{figure}[htbp]
    \centering
    \subfloat[\centering Na\"{i}ve-$\mathrm{N}_{\text{in}}$ MF-EIG ]{{\includegraphics[width=0.49\textwidth]{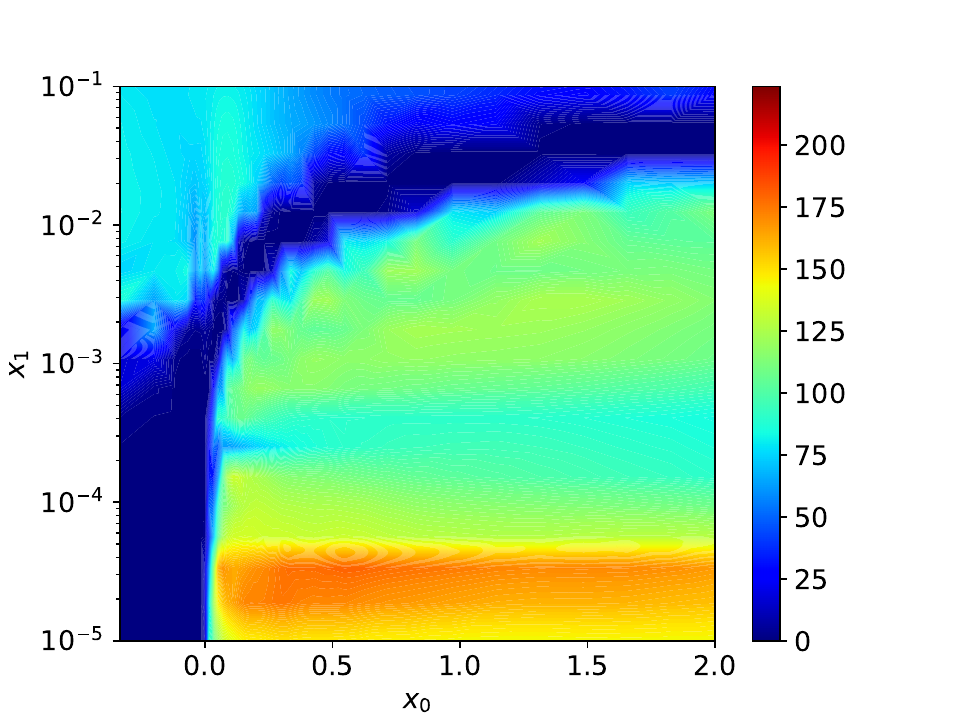} }}
    \subfloat[\centering Optimal-$\mathrm{N}_{\text{in}}$ MF-EIG ]{{\includegraphics[width=0.49\textwidth]{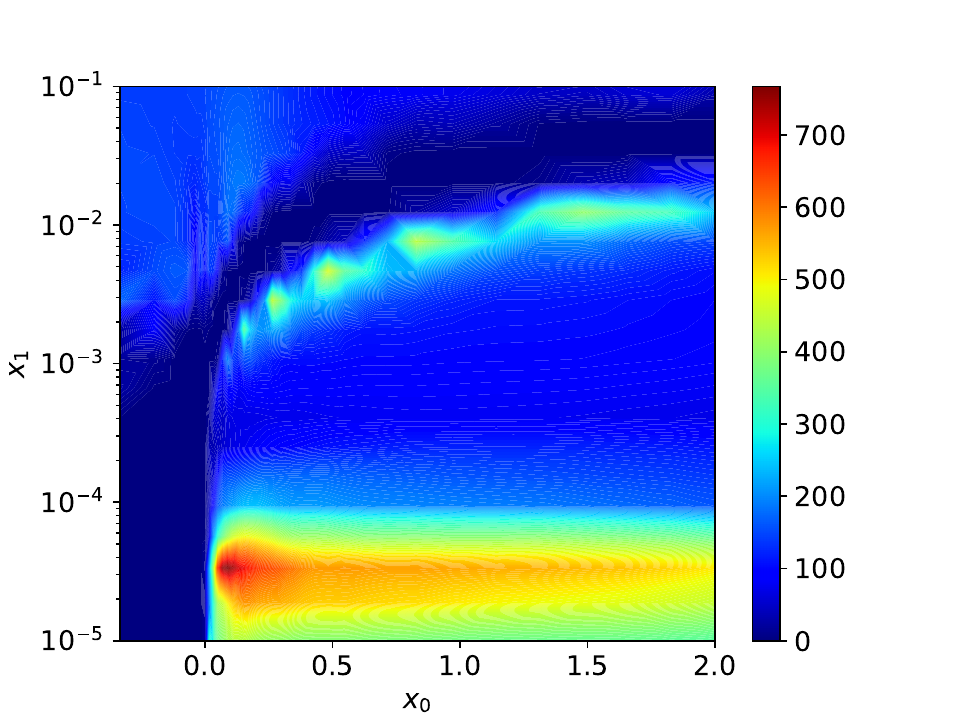} }}
    \caption{Case 2. Empirical variance reduction ratio relative to single-fidelity NMC obtained from 22 repeated estimation trials using the na\"{i}ve-$\mathrm{N}_{\text{in}}$ and optimal-$\mathrm{N}_{\text{in}}$ MF-EIG estimators.
    }
    \label{fig:vrrs}
\end{figure}

Figure \ref{fig:vrrs} plots the variance reduction ratio across the zoomed-in portion of $\Design$. 
The reduction ratio is low in regions far from the flat plate because, as noted earlier, these areas exhibit near-zero EIG and also near-zero variance (i.e., all samples yield roughly the same outcome). 
Consequently, there is little to no estimator variance to reduce in these regions.
At the same time, reducing estimator variance for these low- or non-informative designs would not be of interest anyways. 
In contrast, 
MF-EIG performs incredibly well at the optimal design location, just off the leading edge of the flat plate but well below the apparent upper edge of the boundary layer, 
with the na\"{i}ve-$\mathrm{N}_{\text{in}}$ and optimal-$\mathrm{N}_{\text{in}}$ MF-EIG estimators achieving 94 and 233 times variance reduction, respectively, relative to single-fidelity NMC. The MF-EIG estimators also produce considerable variance reduction around the apparent boundary layer, but the effects are less drastic. 

\begin{figure}[htbp]
    \centering
    \includegraphics[width=10cm]{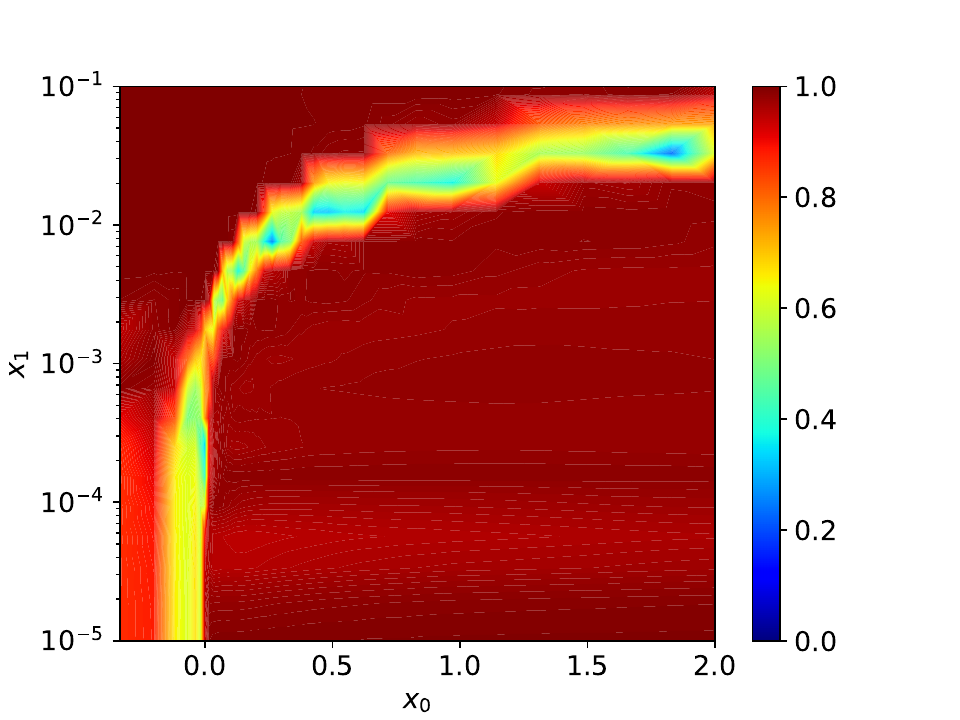}
    \caption{Case 2. Correlation coefficient between each $u_m$ and $u_0$ averaged over $m=1,2,3$.}
    \label{fig:corrs_contour}
\end{figure}

The variation of MF-EIG variance reduction ratio across $\Design$ is likely affected by two main factors. 
The first factor is the inherent design-variation of utility model correlations. 
Figure \ref{fig:corrs_contour} shows the correlation coefficient between each $u_m$ and $u_0$ averaged over $m=1,2,3$. 
Very high correlation emerges in certain regions, such as just off the leading edge of the flat plate. Very low correlation takes place in other regions, such as just above the apparent edge of the boundary layer. These differences are driven by the underlying physical structures and the ability of solvers in capturing them. For example, far-field regions where flows are mostly free-stream without complex fluctuations can be easily captured by all solvers and therefore exhibit high correlation. While one generally expects high correlation to translate to high variance reduction, the high correlation in the free-stream regions do not, as seen in Figure \ref{fig:vrrs}.
This is because their estimator variance is also near-zero and cannot be further reduced, as previously discussed. 
The second factor is the design-averaging process of the covariance matrix. Since the MF-EIG estimators are constructed based on a single covariance matrix averaged over, in this case, the entire $\Design$, the MF-EIG hyperparameters are necessarily suboptimal in regions where the correlation deviates from the average value. 

\begin{figure}[htbp]
    \centering
    \subfloat[\centering Single-fidelity NMC]{{\includegraphics[width=0.49\textwidth]{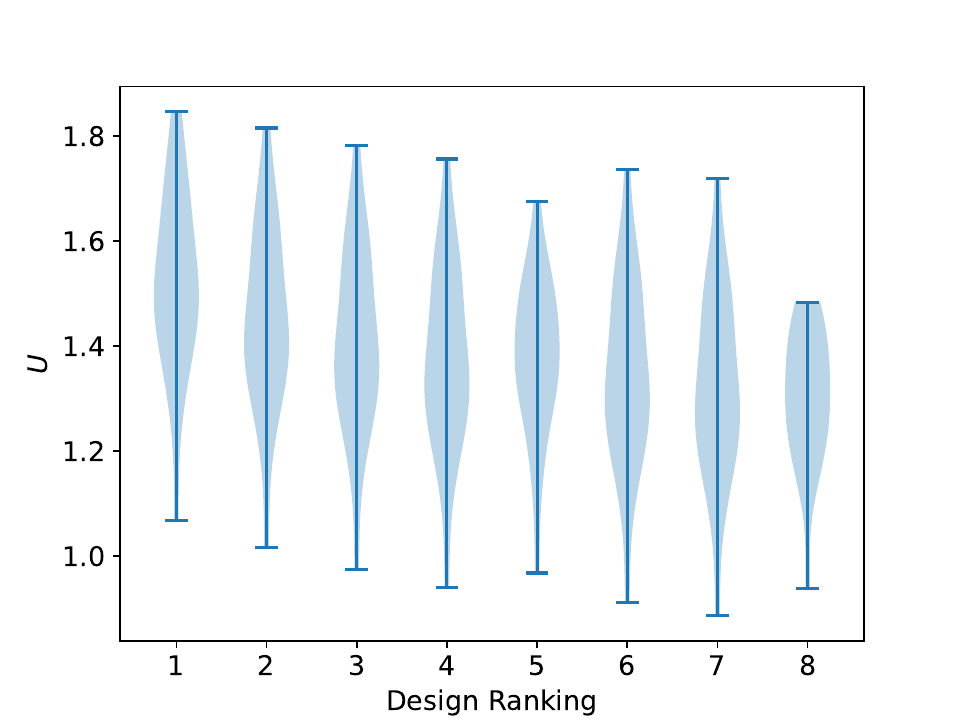} }}\\
    \subfloat[\centering Na\"{i}ve-$\mathrm{N}_{\text{in}}$ MF-EIG]{{\includegraphics[width=0.49\textwidth]{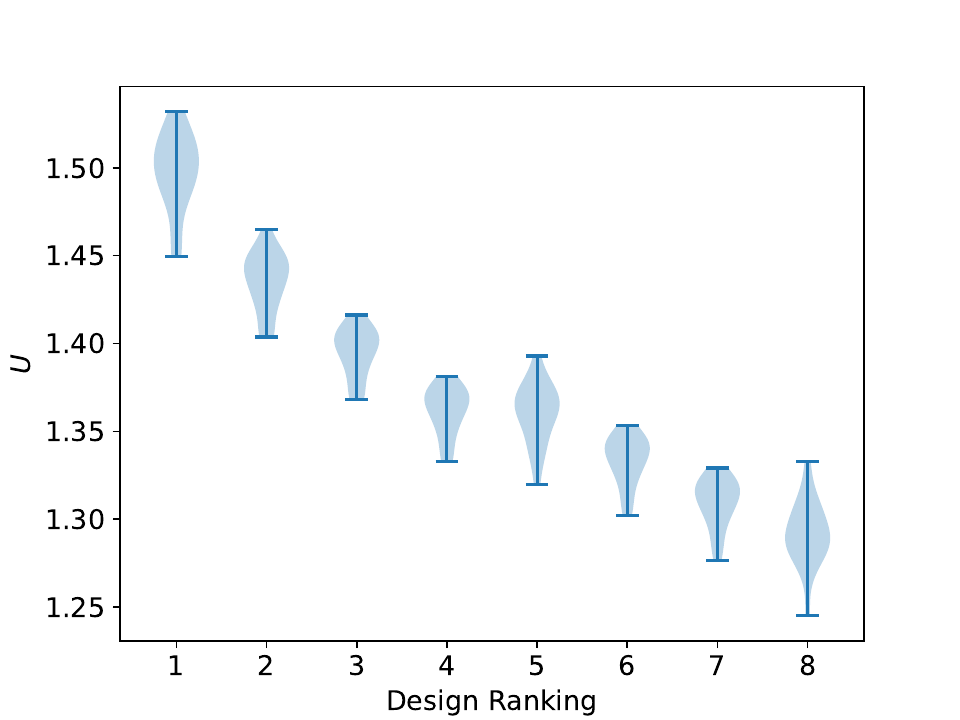} }}
    \subfloat[\centering Optimal-$\mathrm{N}_{\text{in}}$ MF-EIG]{{\includegraphics[width=0.49\textwidth]{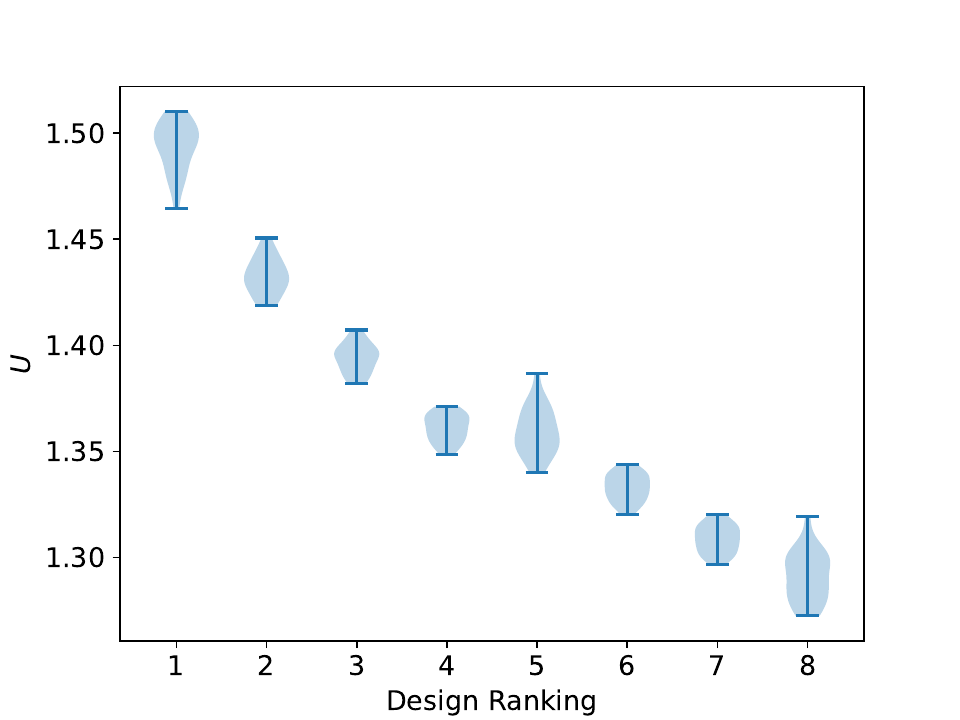} }}
    \caption{Case 2. 
    Distribution of EIG estimates for the 8 most informative designs
    using different estimators. Designs are ranked by the empirical estimator mean using the optimal-$\mathrm{N}_{\text{in}}$ MF-EIG estimator.
    }
    \label{fig:violins}%
\end{figure}

Lastly, we consider MF-EIG in the context of discriminating designs, under the noise of MC estimation, for the purposes of finding the optimal design in \eqref{optimProb}. 
In this case, since the design space $\Design$ is on the computational domain itself and each flow solve computes $g_m$ simultaneously across the entire $\Design$, the EIG evaluations at multiple designs can be obtained for free if we accept to use the same $z$ samples for them---that is, the same flow solves used to estimating EIG at one design can be used for estimating EIG at any other design, without needing any new flow solves.  
This is the same common random number technique mentioned earlier that improves estimator correlation across $\Design$, which aids design comparison.
Even with these considerations, discriminating which design is better remains difficult for single-fidelity NMC, where the estimator variance is large relative to the difference between mean values of competing designs, as seen in Figure \ref{fig:violins}. 
In contrast, the variance reduction provided by MF-EIG greatly alleviates this issue, now having very little overlapping of the EIG distribution between designs. Consequently, the optimal design emerges more distinctly as the best choice among the available options.

\section{Conclusions}
\label{sec:conclusions}

This paper introduced MF-EIG, a novel EIG estimator for OED problems that accommodates general multi-fidelity ensembles of models under the ACV framework. 
Notably, this estimator is unbiased with respect to the high-fidelity mean, and minimizes variance under a given computational budget.
We achieved this by first presenting a reparametrization of the EIG that formulates its expectations to be independent of the data models, a requirement for compatibility with the ACV framework.
This reparameterization also provided insights into the inner mechanics of EIG estimation and the notion of data model fidelity. 
We then provided specific
examples of NMC-style utility models under additive independent and scaled data noise forms, and proposed
practical enhancements via inner-loop sample size optimization and sample reuse techniques.
We demonstrated the MF-EIG estimator in two numerical examples: a one-dimensional nonlinear benchmark where we made extensive comparisons under different settings, and
a multi-dimensional turbulent flow
problem involving the calibration of SST turbulence closure model parameters within the RANS model where we also provided explanations to the behavior of results through physical principles and insights. 
We validated the estimator's unbiasedness and variance reduction through these
examples, where MF-EIG achieved one- to two-orders-of-magnitude variance reductions, with moderate additional improvements from optimizing the inner-loop samples and substantial improvement from reusing the inner-loop samples. 
Based on these results, we strongly recommend utilizing inner-loop sample reuse. Where practical, we also recommend optimizing for low-fidelity inner-loop samples sizes, though its impact appears to be less pronounced than that of sample reuse.
We implemented MF-EIG using the MXMCPy toolbox, and made our code available at: \url{https://github.com/tcoonsUM/mf-eig/}.

While we demonstrated the effectiveness of the MF-EIG estimator, it also has limitations. 
One limitation, in fact a general requirement for all successful multi-fidelity methods, is that there must exist lower-fidelity models that can be exploited, especially those that can produce outputs highly correlated with the high-fidelity model while needing significant less computational costs. 
Interestingly, in the MF-EIG case, such low-fidelity models may be systematically ``manufactured,'' for example by restricting the number of inner-loop samples in an NMC-style utility model. As we focused on situations where the fidelity is controlled by the underlying physical model, such techniques are left for future work. 
Another limitation is that the low-fidelity models must share the same parameterization as the high-fidelity model---i.e., they must have the same $Z=\{\Noise,\Theta\}$. An interesting avenue thus entails accommodating scenarios where the models share only a subset of parameters, for example through adaptive basis sampling methods~\cite{zeng_multifidelity_2023}. 

Another major future direction is to incorporate the MF-EIG estimator within the OED optimization problem in~\eqref{optimProb}, where opportunities exist to further enhance the MF-EIG efficiency. First, the estimator structure does not need to be static across the design domain. As seen in our numerical examples, the correlation among the utility models can vary significantly across the design space. When a single estimator structure is used across the entire design space, its performance will necessarily become suboptimal in certain regions. Furthermore, as the ultimate goal is to find the optimal design, the EIG  at some designs may not need to be well-estimated, while at other designs it might be critical. Adapting the MF-EIG estimator structure in a more goal-oriented manner, to better leverage its role within the overall OED problem or optimization under uncertainty in general, could bring substantial benefits. 
Second, the often-expensive step of pilot sampling for estimating the utility model correlation can be also improved, driven by questions such as: What are the effects of pilot sample size and how many pilot samples are needed? How can correlation information be shared across designs in a principled manner, so that the total cost of pilot sampling can be reduced? When is it worthwhile to take another pilot sample, versus saving its computational resource for enhancing the MF-EIG estimator or furthering the OED optimization steps?
Answers to these questions will require substantial future investigations into the uncertainty quantification of correlation estimation, modeling and extrapolating correlation across designs, and understanding the role and importance of correlation error towards identifying the optimal experimental design.

\section*{Acknowledgments}
This material is based upon work supported in part by the National Science Foundation Graduate Research Fellowship under Grant No. DGE 1841052.
This research is based upon work supported in part by the U.S. Department of Energy, Office of Science, Office of Advanced Scientific Computing Research, under Award Number DE-SC0021397.  
This work relates to Department of Navy award N00014-23-1-2735 issued by
the Office of Naval Research.
This research is supported in part through computational resources and services provided by Advanced Research Computing at the University of Michigan, Ann Arbor.

\bibliographystyle{siamplain}
\bibliography{references}

\end{document}

%% file: ex_shared.tex
\usepackage{lipsum}
\usepackage{amsfonts}
\usepackage{graphicx}
\usepackage{epstopdf}
\usepackage{algorithmic}
\usepackage{amssymb}
\usepackage{amsmath}
\usepackage{array}
\usepackage[english]{babel}
\usepackage{hyperref}
\usepackage{xcolor}
\usepackage{subfig}
\usepackage{float}
\ifpdf
  \DeclareGraphicsExtensions{.eps,.pdf,.png,.jpg}
\else
  \DeclareGraphicsExtensions{.eps}
\fi

\newsiamremark{remark}{Remark}
\newsiamremark{hypothesis}{Hypothesis}
\crefname{hypothesis}{Hypothesis}{Hypotheses}
\newsiamthm{claim}{Claim}

\headers{A Multi-fidelity Estimator of the EIG for BOED}{Thomas E. Coons, and Xun Huan}

\title{A Multi-fidelity Estimator of the Expected Information Gain for Bayesian Optimal Experimental Design\thanks{Submitted to the editors February 5, 2025.
}}

\author{Thomas E. Coons\thanks{Department of Mechanical Engineering, University of Michigan, Ann Arbor, MI, 48109
  (\email{tcoons@umich.edu}, \url{https://sites.google.com/umich.edu/tcoons/home}).}
\and Xun Huan\thanks{Department of Mechanical Engineering, University of Michigan, Ann Arbor, MI, 48109 
  (\url{https://uq.engin.umich.edu/}).}
  }

\usepackage{amsopn}

%% file: macros.tex
\definecolor{darkred}{rgb}{.7,0,0}

\definecolor{darkgreen}{rgb}{.15,.55,0}

\definecolor{darkblue}{rgb}{0,0,0.7}

\newcommand{\design}{\xi}
\newcommand{\Design}{\Xi}

\newcommand{\noise}{\epsilon}
\newcommand{\Noise}{\mathcal{E}}

\newcommand{\Var}{\mathbb{V}\text{ar}}
\newcommand{\Cov}{\mathbb{C}\text{ov}}

\newcommand{\argmax}{\operatornamewithlimits{argmax}}

\newcommand{\EE}{\mathbb{E}}

